# Physics-informed time-series forecasting of perovskite photoluminescence stability

Alexander Wieczorek[1,#], Manuel Kober-Czerny[1,#,*], Fábio Lopes[1,#], Austin George Kuba[2], Leon Müller[3], Christian M. Wolff[2], Jason Hattrick-Simpers[4,5], Sebastian Siol[1,*]

[1] Laboratory for Surface Science and Coating Technologies, Empa–Swiss Federal Laboratories for Materials Science and Technology, Ueberlandstrasse 129, Dübendorf CH-8600, Switzerland.

[2] Institute of Electrical and Microengineering (IEM), Photovoltaic and Thin-Film Electronics Laboratory, EPFL –École Polytechnique Fédérale de Lausanne, Switzerland.

[3] Fritz-Haber-Institut der Max-Planck-Gesellschaft, Faradayweg 4-6, D-14195 Berlin, Germany

[4] Department of Materials Science and Engineering, University of Toronto, M5S 3E4, Toronto, ON, Canada.

[5] Acceleration Consortium, University of Toronto, M7A 2S4, Toronto, ON, Canada.

[#] These Authors contributed equally

Corresponding Authors: Manuel.Kober-Czerny@empa.ch, Sebastian.Siol@empa.ch

## ABSTRACT

Accelerated ageing using elevated temperatures and illumination is one of the most common methods to rapidly study the stability of novel semiconductor materials. However, as the pace of materials discovery continues to accelerate, even faster stability evaluations are needed. A physics-informed time-series forecasting algorithm designed to predict the long-term photoluminescence stability of metal halide perovskites is presented. A diverse experimental dataset of 167 metal halide perovskites is collected, including different crystallinities and compositions. These are stressed using heat and light, while the photoluminescence (PL) is monitored. The >86k collected PL spectra are featurized using a physics-informed model, and a hybrid CNN-LSTM model is trained to forecast the PL intensity during degradation of samples unseen during model training. Notably, the approach generalizes across the material groups and outperforms baseline benchmarks. Furthermore, the physics-based featurization ensures explainability, enabling analysis to identify critical stability descriptors for given predictions. It is expected that this approach will be adapted to other types of time-series data and enables a pathway to significantly reduce experimental testing times.

# I. INTRODUCTION

The development of new semiconductors is fundamental to enabling advances in optoelectronic devices. This ranges from new detectors suitable for medical applications[1], to photovoltaic cells for low-cost green energy[2], and light emitting materials for displays[3]. To this end, new chemistries are investigated in research labs across the globe. Beyond pristine properties, validation of long-term stability marks a major milestone towards the widespread application of new materials.

Metal halide perovskites (MHPs) have been at the frontier of this research for more than a decade, with outstanding demonstrations for a range of optoelectronic applications[4], [5], [6]. Despite performances often surpassing those of their established counterparts based on Si, applications are thus far mostly limited to non-industrial cases[7]. This is in large parts due to their low intrinsic stability under operating conditions[8]. Consequently, improving stability is one of the most critical challenges in the development of these materials.

As the stability requirements, e.g. for photovoltaic applications, are on the order of 20 years[9], multiple strategies have been employed to determine the ageing behaviour more quickly. Most prominently, testing under increased stress conditions, e.g. elevated temperatures or humidity, is used[10]. However, resulting test cycles may still be on the order of months and degradation mechanisms at higher temperatures may not be present at typical operating conditions. Over the past decade an increasing number of research groups have adopted data-driven, high-throughput screening strategies for materials discovery and fabrication. By integrating automation, advanced characterization, and machine-learning models, these workflows prioritize and execute experiments more efficiently, enabling order-of-magnitude accelerations in the optimization of functional materials. Yet, despite their proven impact on materials development, high-throughput approaches remain largely underexploited for rapid stability testing[11]. As previously highlighted by our group[12], parallel testing may indeed result in a higher throughput, unlocking large experimental data sets of altered materials properties under degradation simultaneously. However, for highly optimized materials, long test cycles may still mark a major bottleneck for fast iterative improvement of the materials based on stability data.

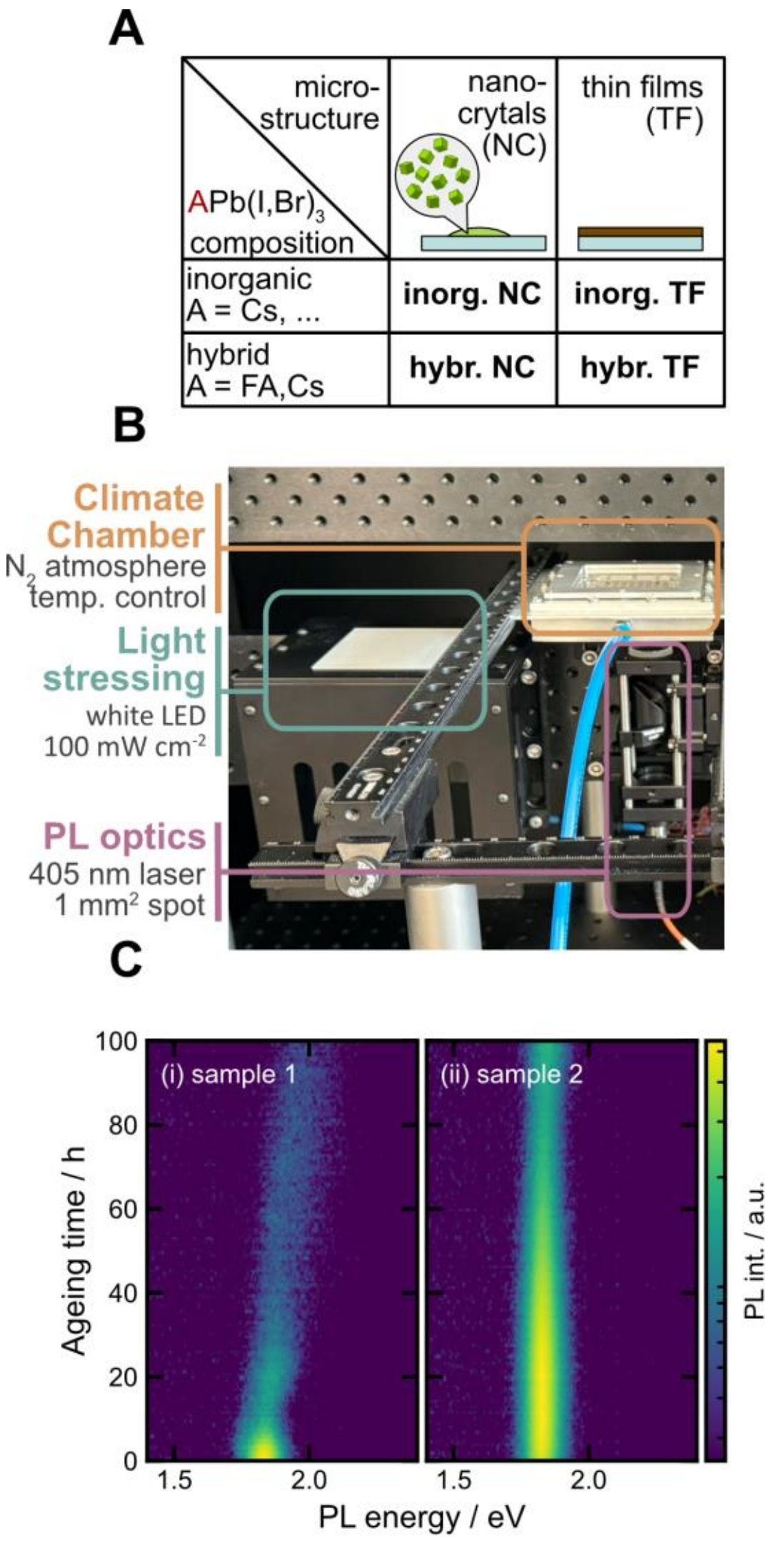

**Figure 1 A**: Overview of the different metal halide perovskite chemistries and crystallinities studied in this work. **B**: Picture of the setup used for the high throughput ageing of the perovskite samples. They were placed inside the climate chamber, where the atmosphere and temperature could be controlled. A 405 nm laser was used to excite the samples and obtain photoluminescence spectra. **C**: Representative time-dependent PL spectra acquired for two individual samples under continuous degradation.

Time-series forecasting algorithms trained on experimental data sets may offer a clear pathway to reduce testing time while maintaining stress conditions close to those found in applications. Here, historical degradation data can be used to train a model which predicts long-term trends from short-term data. Similar approaches[13] have been most recently explored for battery stability predictions. While such methodology is in principle applicable to any semiconductor system, including organic photovoltaics, kesterites, perovskites, and III-V systems, data-driven approaches remain broadly underutilized. In the case of perovskite semiconductors, existing efforts are limited to edge cases for highly specific chemistries[14], [15]. This is largely due to the various degradation pathways that can occur in these materials, making predictions challenging[16], [17]. Fully data-driven approaches may help to overcome this, but often come at the expense of explainability, thereby limiting trust and usefulness in these models and their predictions.

In this work, we aim to address this gap by employing a combination of physics-informed featurization of experimental data and time-series forecasting to accelerate the research of metal halide perovskite (MHP) material stability. As shown schematically in **Figure 1A**, we fabricated a series of samples spanning four different material groups which include different compositions and crystallinities. The samples were monitored during degradation using photoluminescence (PL) measurements in a custom-built high-throughput platform[12] designed for the automated optical characterization of materials libraries, each containing many unique samples. We deliberately focused on PL properties in this study as it is closely linked to properties relevant to various optoelectronic device applications. To bridge the gap between experimental data and forecasting, we developed an automated spectral segmentation algorithm to extract physically meaningful features from photoluminescence (PL) spectra during degradation. We present a forecasting model based on a combined convolutional neural network (CNN) and long-short-term-memory (LSTM) architecture, which clearly outperforms naïve predictions especially over longer input times. The physics-informed approach results in an intrinsically interpretable model that allows to use SHapley Additive exPlanations (SHAP) to identify the most important physical parameters underpinning PL yield stability.

## II. METHODS

### A. The Dataset

For this work, we fabricated MHP samples with a range of compositions and crystallinities to represent the commonly studied materials space: this includes films of inorganic nanocrystals (inorg. NC), polycrystalline inorganic thin films (inorg. TF), films of organic-inorganic hybrid nanocrystals (hybr. NC), as well as polycrystalline organic-inorganic hybrid thin films (hybr. TF). All details about the material fabrication can be found in the supporting information. For each individual experiment, up to 45 samples (compositions) of each group were deposited on the same glass substrate in a grid (here: "*library*"; see inset of Figure S1 for clarification). We used a custom-built high-throughput platform[12] as shown in **Figure 1B** to automatically stress and monitor all samples on a materials library simultaneously. All samples were stressed with a white light LED (100 mW $cm^{-2}$) and heat under constant nitrogen flow. For the heat and atmosphere control, the libraries were placed inside a climate chamber[12]. We chose stress temperatures expected during operation of corresponding optoelectronic devices: 25 or 60 °C for nanocrystalline samples and 85 °C for polycrystalline ones (see Table S1 for more details).

Approximately every 20 min. a PL spectrum was automatically measured for each sample to track the degradation. The samples were aged between 6 and 19 days in total and we obtained nearly 86k PL spectra (see Table S1 for more details). In **Figure 1C**, we show the PL traces of two individual samples as a function of ageing time to demonstrate the variance of PL behaviour during degradation. At first, a shift in the emission maximum is visibly due to the different composition. More interestingly, we also observe a difference in the PL intensity over time. The PL intensity in MHPs is strongly affected by defect densities in the material and can be linked to optoelectronic device properties, such as $V_{oc}$ in solar cells or brightness of light-emitting diodes[18], [19], [20], [21], [22]. More directly, PL intensity is the metric most relevant to commercially available quantum dot displays, in which a high-energy light source excites a photoluminescent NC layer[23]. For this work, we hence take the normalized area of the PL spectrum (here: *Norm area*) as a measure of stability of our perovskite compositions during ageing.

It is also the property that we will predict using our forecasting model later. In **Figure 2A**, we show *Norm area* over time for the four different MHP groups. The colour of the data corresponds to the sample position on the respective libraries (see Figure S1). Overall, we observe quite complex behaviour, including increases and decreases of the PL area over time with non-linear trends.

We want to highlight the behaviour of the inorg. TF samples at this point. While most of the samples behave similarly, there are a handful that are different in two regards: a) a single sample appeared to increase in normalized PL and then stabilize, while b) a group of samples show an initial drop in normalized PL followed by a significant increase and a subsequent drop. To show that this is a real behaviour of the material system, rather than a measurement artifact, we show the time-dependent PL spectra of all material groups in Figure S2. For the inorg. TF case we then also show representative samples for both exceptions. We also show the bromide and cesium fractions, as well as the thickness as obtained from X-Ray fluorescence (XRF) alongside the initial PL intensity ($t$ = 0) in Figure S3. The sample representing category (a) exhibited the lowest initial PL intensity, the highest bromide and cesium fractions, and is also the thinnest sample on this library. During degradation the increase in PL intensity is accompanied by a blue-shift, which may be due to loss of iodine from the material, as previously reported for nanocrystals of the same chemistry[24]. For the second category (b), we find a smaller blue shift that accompanies the increase in PL intensity. Further, all samples of this type showed generally lower bromide and cesium fractions and also tend to have low initial PL intensities. We hypothesize that these samples are not only outliers with regards to their stability behaviour but represent "extreme" cases in terms of composition and thickness among the ones investigated. As a result, we conclude that both behaviours are real in these perovskite compositions and should hence be part of the dataset, reserving an opening to a composition-thickness-stability regime not present in the other samples.

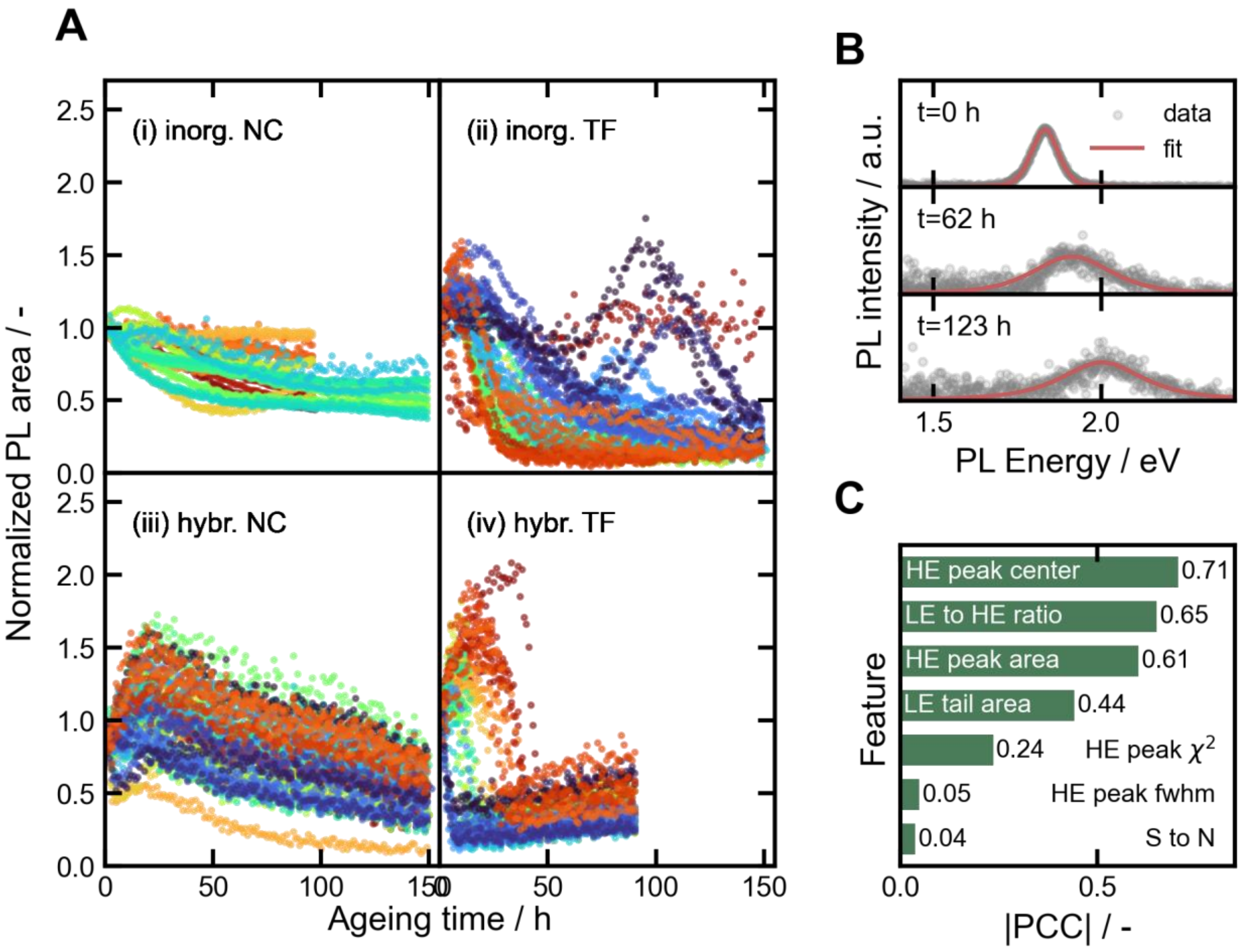

**Figure 2 A**: The normalized PL area (Norm area) is shown for the 4 different material groups of samples used in this study ((i)-(iv). **B**: Results for automated segmentation of PL features into a high energy (HE) and low energy (LE) feature at various points in time. **C**: Absolute Pearson correlation coefficients (PCC) calculated for ΔNorm area at $t$ = 80 h vs. the features at $t$ = 0 h across all material groups and samples.

To extract physically meaningful insights from our spectra, we developed an automated spectral segmentation algorithm (HE+LE model). In short, the high-energy (HE) side of the PL peak was fitted to a symmetric Gauss-Lorentz pseudo-Voigt peak shape. The residual area of the peak is ascribed to a low energy (LE) feature. The full spectral fitting routine is explained in the supporting information and the code used provided alongside this manuscript. A list of all the 8 extracted features is given in Table S2 and equivalent figures to Figure 2A for all features are shown in Figures S3-6. **Figure 2B** depicts the automated fits for a representative sample at $t = 0$ h, 62 h, and 123 h. All features were then taken as inputs for our time-series forecasting model, while *Norm area* was used as the output for the predictions.

While time-series forecasting is a powerful tool, a lot can already be learned from data exploration of this comprehensive dataset. For instance, we aimed to understand, if there are features of the PL spectrum, which are crucial for improved long-term stability. Here we use the change of *Norm area* as a measure for long-term stability: For a perfectly stable material, we do not expect any change in PL spectrum, hence no change in *Norm area* over time. We then check whether any of the 8 features at $t = 0$ h are correlated with the change in *Norm area* at $t = 80$ h. The absolute Pearson correlation coefficients (PCC) are shown in **Figure 2C** for the respective features. It becomes apparent that the stability is mainly affected by the centre energy of the PL peak (HE peak center), which in MHPs encodes both the composition and the crystallinity. It is worth noting that the NC were aged at different temperatures than the TF and consequently show better long-term stability. Since the NC and TF have different peak centres as extracted from the HE side, this may enhance the correlation with the HE peak center as well. However, regardless of this, we also find strong influence from features related to the PL area (HE peak area), and the LE feature (LE to HE ratio and LE tail area). We will later be able to understand the time-series forecasting model better by taking these correlations into account.

### B. Data Preparation Pipeline

For this work, we *pre*-treat the data to be able to extract physically relevant features from the PL spectra. In **Figure 3A** we show schematically the different steps that are taken from the raw PL data and the final data frame of time-resolved features. In short, we first remove both the substrate background as well as the remaining baseline we obtain as a linear fit between the edges of the cropped

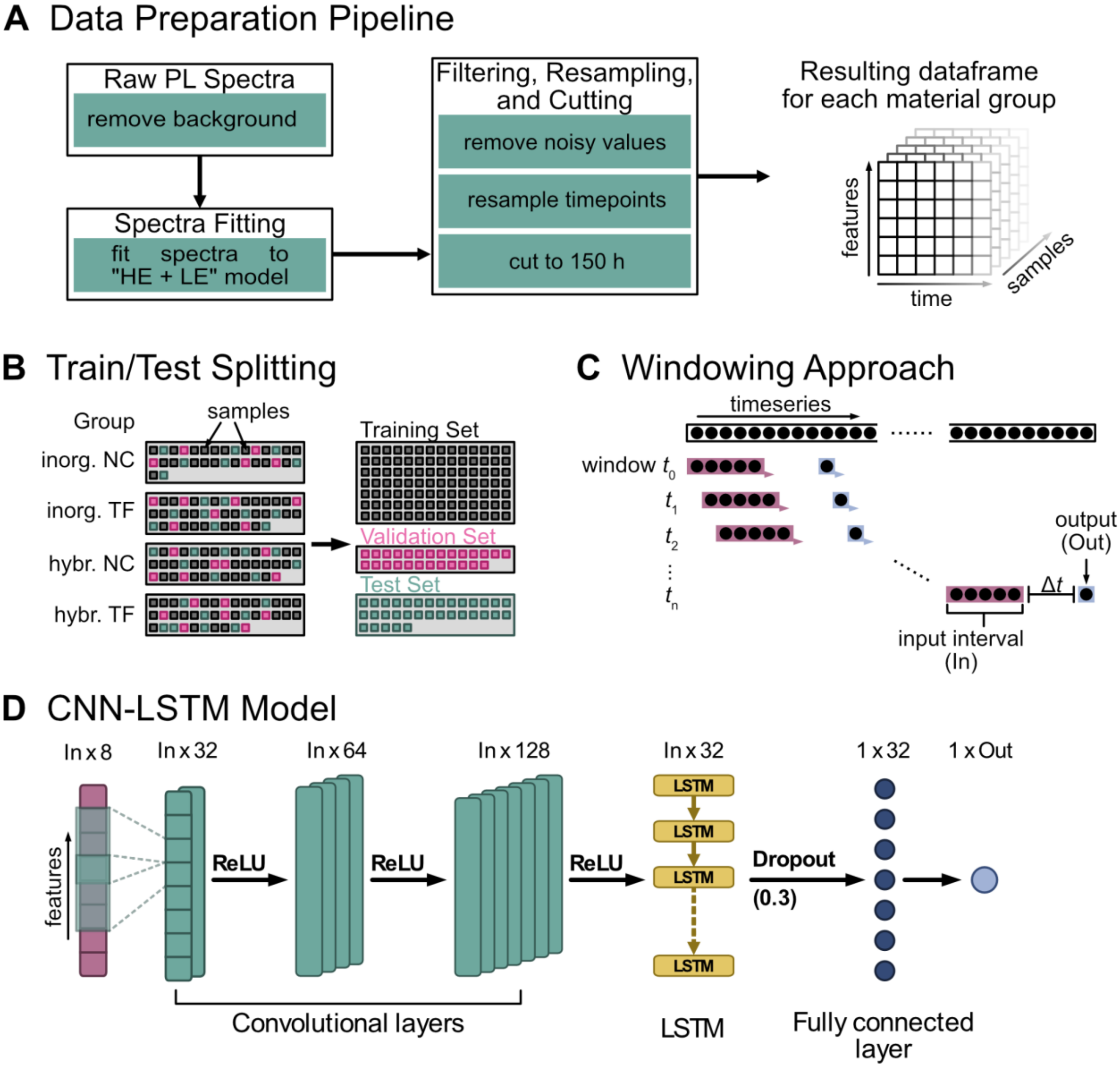

**Figure 3 A**: An overview of the data preparation pipeline from raw photoluminescence data to a data frame of the features that can be used for the training of the time-series forecasting model is shown. **B**: The feature data frames are split into the training, testing, and validation set. Here, we show that the test set is split off by sample from each material group. The remaining samples of all material groups are mixed and split 8:2 by sample again into the training and validation sets. **C**: Here the sliding window approach is shown: The first window ($t_0$) starts at $t$ = 0. The input interval (In) is moved ahead by 1 hour and the output time point (Out) is defined with respect to the end of the input interval via $\Delta t$. **D**: Schematic of the architecture of the CNN-LSTM model. The input is the In x 8 features. First, this is passed through three CNN layers with rectified linear units (ReLU) in between. The result is passed into a layer of LSTM blocks. After 30% of all neurons have been dropped and the result is linearized, a single output (normalized PL area) is obtained.

PL spectra (see supporting info for more details). The resulting spectra are then used to extract parameters using the "HE+LE model" as described before and in the supporting information. The extracted features are then filtered by removing trailing values where no PL peak was present in the spectrum. Each time-series is then re-sampled and forward-filled with 1 h resolution. This is necessary to use the time-series data for forecasting later. All timeseries are cut to 150 h, to make the different material groups more comparable in terms of amount of training data. Lastly, we removed samples (time series) from the datasets, where more than 10% of the timepoints were missing. We found that 10% was a good trade-off between removing corrupted data and maintaining a large number of samples. This resulted in the removal of 4 samples in total (see Table S1). At the end, we are left with approx. 20k timepoints across 163 time series for training, validation and testing of the model (see **Table 1**).

**Table 1** Overview of the amount of data (samples and total timepoints) obtained for each MHP group after the data-preparation has been completed. In addition, the number of samples used for training, validation, and testing is summarized for each group.

| group | total timepoints | samples | for training | for validation | for testing |
|---|---|---|---|---|---|
| inorg. NC | 3'812 | 32 | 20 | 5 | 7 |
| inorg. TF | 5'894 | 42 | 26 | 7 | 9 |
| hybr. NC | 6'750 | 45 | 28 | 8 | 9 |
| hybr. TF | 4'000 | 44 | 28 | 7 | 9 |
| **total** | **20'456** | **163** | **102** | **27** | **34** |

### C. The Time-Series Forecasting Model

We obtain one data set for each material group, as shown in the **Figure 3A**. These are first split 80/20 by sample into train and test sets and the train sets are split again 80/20 by sample into train and validation sets (see **Figure 3B** and Table 1). In Figure 2A we could see that for instance for the inorg. TF group there are a handful of samples that behave significantly differently than the rest of the compositions. This may lead to artifacts during model training, if the train/test/validation split puts too much or too little emphasis on these compositions. We hence repeated the splitting 30 times and, in the results, show the average of all resulting models.

We train the model for a maximum of 500 epochs, for a given set of hyperparameters, to reduce the error on the training data, while monitoring the error on a separate validation set. Early stopping regularization is employed to avoid overfitting. Specifically, if the validation error has not minimized for more than 50 epochs (early stopping patience), the training stops and the model falls back to the one 50 epochs ago. The input of our model (here: input interval) is the interval of hours of historical data used for the prediction. The output (here: $\Delta t$) is a discrete point in the future that is predicted and defined by its distance to the last point of the input interval. We further use a windowing approach (see **Figure 3C**): For a given input interval/$\Delta t$ setting, the first window starts at $t = 0$ h, the second is moved ahead in time by 1 h. From a single time-series we thereby generate multiple input interval/$\Delta t$ combinations that enrich the training, validation, and test datasets. As a result, the model is learning to predict the PL area at the relative temporal distance $\Delta t$ given a certain input interval and not depend on the initial value at $t = 0$ h. This ensures time-invariance, as the model learns based on immediate temporal relationships rather than the absolute timestamp of the measurement.

We used all 8 features as the input and only the normalized PL area (Norm area) as the output. Our time-series forecasting model uses a combination of a convolutional neural network (CNN) and a long-short-term-memory (LSTM) layer (short: CNN-LSTM)[25], [26], [27]. CNN-LSTM can thus be named as a hybrid architecture made of convolutional and LSTM layers (see **Figure 3D**): The convolutional layers extract local patterns of the time-series and also combine the different features used as input data, while the LSTM part can process the time sequence. Rectified linear units (ReLU) are used to connect the CNN layers and the CNN to the LSTM layer to avoid the model collapsing into a linear function. Without the ReLU's, the model would be restricted to linear mappings between the input features and the output (Norm area), imposing an assumption of linearity that may not be justified a priori. A drop-out layer removes 30% of neurons at random during training to reduce the risk of overfitting and improve the model's generalizability. Lastly, the output of the LSTM is passed through a fully connected layer, which maps it to the final prediction value. During the training, we fixed most hyperparameters (learning rate: $5{\cdot}10^{-4}$; batch size: 64; early stopping patience: 50; max. epochs: 500) and instead only varied the input interval and $\Delta t$. The order of windows (i.e. specific combinations of input intervals and $\Delta t$) is randomized between epochs to reduce bias during training.

## III. RESULTS & DISCUSSION

### A. Model training and performance

We first had to ascertain that the CNN-LSTM could be trained on the given data (training and validation loss are shown in Figure S8). After training the model, the testing of our model is done on the test sets, which are compositions that the model has not seen during the training. This is to ensure that the model is somewhat generalizable and can perform well in real-world scenarios. We repeated this process for input intervals and $\Delta t$ of 1, 3, 6, 12, 24, 48, and 72 h, each and show the mean absolute percentage error (MAPE) of the 30 train/test/validation splits in a 7×7 grid **Figure 4A** (for completeness, we show the root-mean-square-error and $R^2$ in Figure S8). Some trends can be picked up within the grid. For instance, models with smaller input intervals and larger $\Delta t$ tend to perform worse during the testing. This makes sense, as the model has less temporal information in short input intervals to learn from. Hence, models with longer input intervals generally perform better during testing. To confirm our forecasting approach is truly effective we compared these results against a naïve benchmark[28], [29]. This benchmark simply takes the last value of the input interval as the prediction for $\Delta t$. This was again repeated for all train/test/validation splits and the difference to the model MAPE was calculated as $\Delta\text{MAPE} = \text{MAPE}_{\text{benchmark}} - \text{MAPE}_{\text{model}}$.

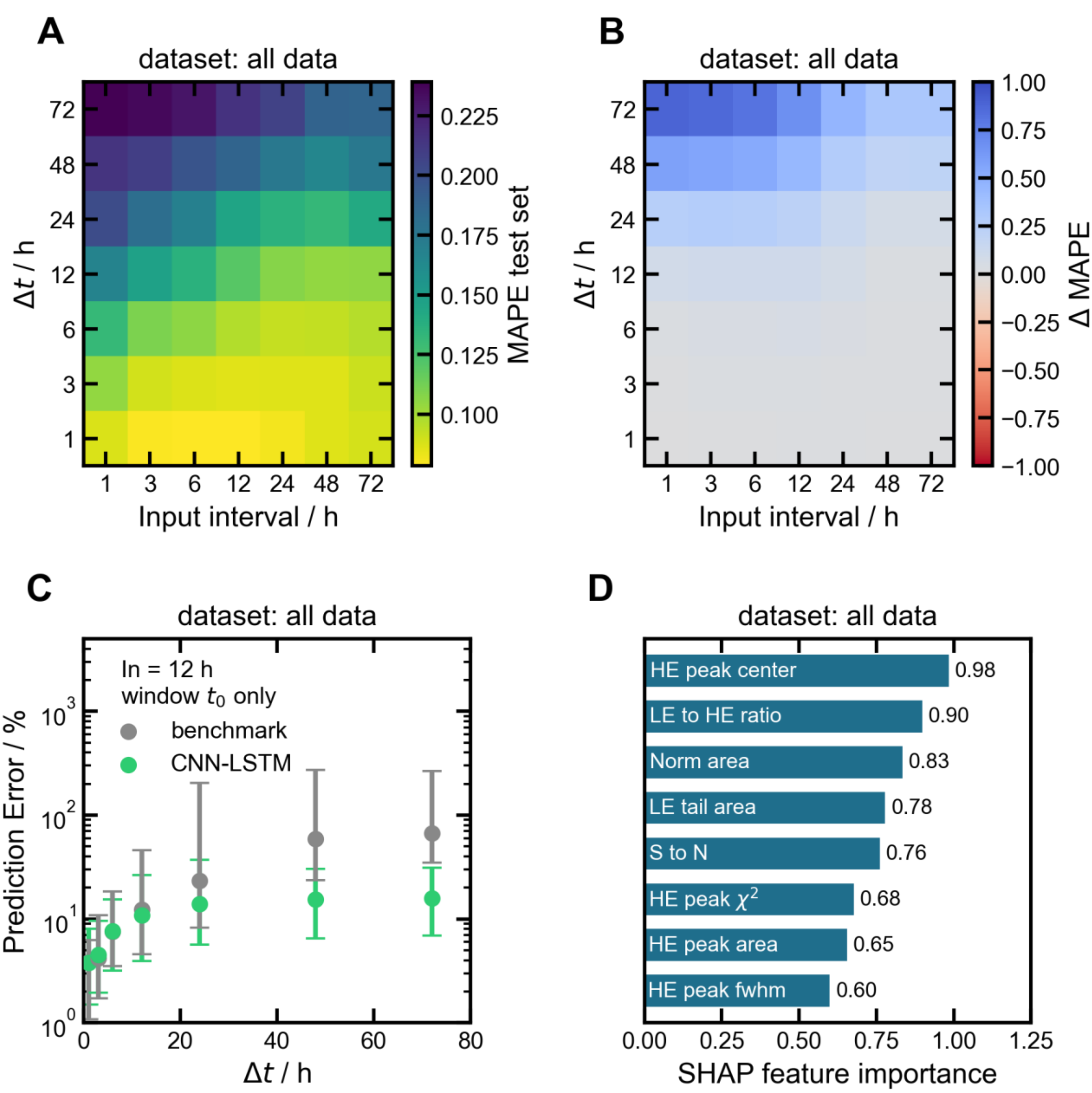

**Figure 4 A**: Mean absolute percentage error (MAPE) of the test sets across all 30 train/test/validation splits and the full 7x7 grids of input interval and $\Delta t$. It is shown for the case where all data was used for training and testing. **B**: The difference in MAPE to the naïve benchmark is shown for the same data as in A. **C**: The prediction error is shown after the initial input window ($t_0$) of 12 h for different $\Delta t$. The naïve benchmark is shown alongside. The points are the median values and the error bars are the interquartile range. **D**: SHAP feature importance for all 8 features used by the model. We show the average of 5 different input interval/$\Delta t$ combinations across the 7x7 grid (3/3, 3/48, 12/12, 48/3, 48/48).

The results are shown in **Figure 4B**. A $\Delta\text{MAPE} > 0$ indicates that the model performs better than the benchmark. This is clearly the case for larger $\Delta t$, where the LSTM component understands the temporal evolution of the extracted features. Indeed, this indicates that time-series forecasting is needed to be able to make longer-term predictions.

We can now describe different regions within the 7×7 grid of input intervals and $\Delta t$'s: For small input intervals and low $\Delta t$, the model performs very well, while the performance slightly decreases for larger input intervals. Neither of these regions are any better than the naïve benchmark, indicating no benefit of the CNN-LSTM approach for short-term predictions. In Figure 2A we can see that *Norm area* does not change much on timescales < 10 h, hence it is expected that the naïve benchmark would perform well for small $\Delta t$. For increased $\Delta t$ the behaviour is inverted, as now a smaller input interval leads to a larger error and vice versa. Crucially, this is the region where the CNN-LSTM approach outperforms the naïve benchmark and which is of much higher interest for materials development.

We also investigated the robustness of the model by shuffling different components of the dataset during training only, while validating and testing on the original data: a) the time axis before windowing, and b) the order of the features (results are summarized in Figure S9). We find that the model is not robust to shuffling the time axis as indicated by a $\Delta\text{MAPE} > 0$, meaning that the temporal change of features is indeed an information the model needs to perform well. Similarly, it is not robust towards shuffling the order of the features, suggesting that the model has no inherent understanding of the physical meaning of the features. Instead, the model only knows about the position of a feature in the data frame.

In a real-world scenario, the time-series forecasting model would be used alongside a degradation study to answer the question of how stable a new composition may be. For this, we chose a single input interval (here: 12 h as the middle of the grid) and showed how the prediction error evolved as a function of $\Delta t$. To be realistic, we only focus on the first window of the time-series, which starts at $t = 0$. The result can be seen in **Figure 4C** as the median (and interquartile range) of all 30 train/test/validation splits and across all samples used. Even for longer prediction horizons of 48 h or 72 h the prediction error is never larger than 30% and the median is consistently below 20%. This gives us high confidence in the applicability of our approach to real-world scenarios.

As our model uses real-world physical features as inputs, its predictions are easier to understand. We used SHAP [21] to find out exactly which of these physical features the model relied on most. Here, each combination in the 7×7 grid is its own model and may hence give a different feature importance. We probe the models at input intervals/$\Delta t$ of 3/3, 3/48, 12/12, 48/3, and 48/48 to represent the full grid (see Figure S10). It is striking that all probed SHAP feature importances are almost identical, with minor differences in the less important features. In **Figure 4D** we show the average SHAP feature importance of our 5 probed combinations to represent the entire 7×7 grid. It becomes immediately apparent that *Norm area* is not the most important feature, even though it is the feature to be predicted. Instead, the model relies strongly on the peak energy of the PL peak (HE peak center), which encodes both composition and microstructure of the individual samples. It is also the feature that was the most important indicator for 'stability' as was shown in Figure 2C. Among the four most important features are the ones related to the low-energy PL feature (LE to HE ratio and LE tail area). This finding agrees with literature, where a low-energy tail of the PL emission has been reported to have many underlying mechanisms: In nanocrystals, it may stem from emission from lower-energy surface states[30], [31] or biexcitons[32], while in thin films, low-energy tails have been linked to defect states near the band edge [33], inhomogeneities in halide composition [34], and halide segregation [35], [36]. This finding demonstrates that our models' working principle is linked to fundamental physical principles, resulting in high confidence in our approach for trustworthy predictions.

**B. Learning Transfer in the Metal Halide Perovskite Material Class**

The approach shown in this work can only be useful to the scientific community, if it can be used to predict the degradation of new, unseen materials. In the case presented here it could be novel perovskite compositions or microstructures. To simulate this scenario, we tested whether there was any learning transfer if the model was trained only on a subset of the material groups. We hence used each of the individual material groups, as well as combined sets of the inorganic (inorg.), hybrid (hybr.), nanocrystals (NCs) or thin films (TFs) for training, validation, and testing. In Figures S10-S12 we show the resulting MAPE, RMSE, and $R^2$ on the same 7×7 grids (again after 30 train/test/validation splits) after training and testing each subgroup. It becomes apparent that there are some differences between the grids. For instance, when the model is trained on the inorg. TF group or any group that contains this dataset (inorg. and TFs) it does not perform well during testing, especially for larger $\Delta t$. In contrast, any group containing nanocrystalline data sets (inorg. NC, hybr. NC, NCs) perform much better during testing overall. We understand this behaviour because of the underlying data. In Figure 2A we showed *Norm area* as a function of time for all four material groups. We can clearly see that both NC groups follow a more predictable decay behaviour, and all degrade in a similar way. In contrast, the TF groups show a wider range of degradation behaviours that may make predictions more difficult to make. The inorg. TF group is the only one that has some clear "extreme" cases as aforementioned, which appear after approx. 100 h. These cases disproportionately affect predictions at larger $\Delta$t, and the difficulty in predicting them suggests the presence of degradation regimes not well captured by the model — potentially pointing to distinct underlying mechanisms that warrant further investigation.

To clarify this point more, we can calculate the average (avg) and standard deviation (std) of the MAPE across the 30 splits and 7×7 grids. The av of MAPE can be used as a proxy for the performance during testing, while the std of MAPE can be used to measure the similarity of data sets within the 30 splits. We then compare this to some statistical moments of the different data sets, such as mean, variance, skew, and kurtosis and show the results in Table S3 and Figure S14. We find very strong correlations (PCC > 0.7) of the av. of MAPE with the mean and variance of Norm area, and a strong correlation (> 0.5) with skew. They can be used to understand how the model uses the data. A lower variance indicates that all the data in a data set behave in a similar way, hence when testing on new samples the model assumes they stem from the same narrow distribution, which is easy to predict. A larger mean *Norm area* is equivalent to more stable materials, which in our case are the inorg. NC and hybr. NC. Datasets of more stable perovskite seem to be easier to predict, but they also are the datasets, which have lower variance, so the two are strongly correlated as well. Lastly, a skew closer to zero describes a more normal distribution and leads to a lower av. of MAPE, due to less 'extreme' cases. We do not find any significant correlation of the statistical models of *Norm area* and std. of MAPE.

We then went one step further and tested each model on out-of-distribution data. For example, the model trained on the inorg. TF samples was then tested on the data of all the other subgroups. We then again calculated the av. of MAPE to see how the different models would perform. In **Figure 5A** we show the resulting matrix of data sets we trained the model with ("trained with") and those, we tested the models with ("tested with").

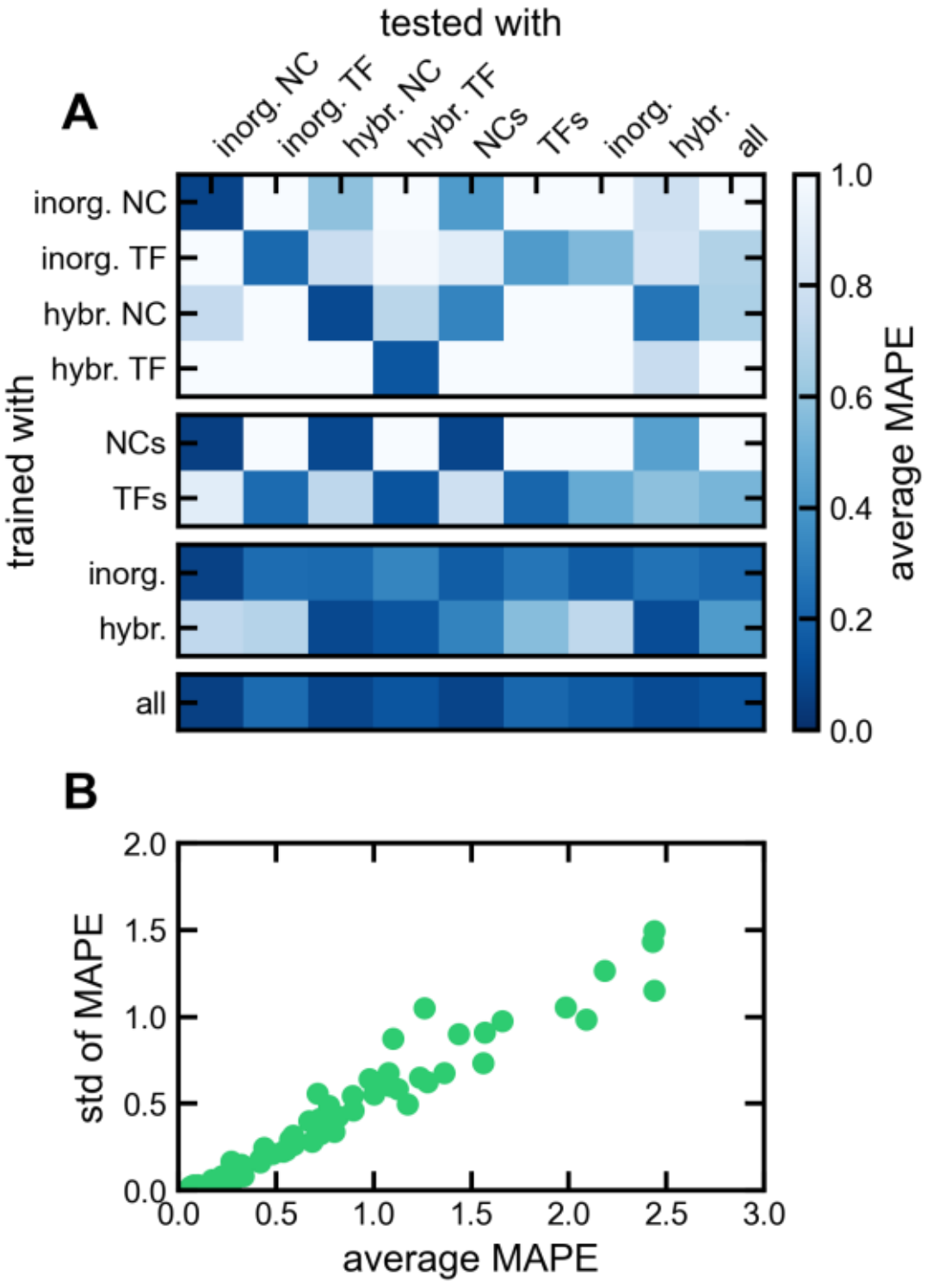

**Figure 5 A**: Average mean absolute percentage error (MAPE) of the 30 train/test/validation splits and across the full 6x6 grids of input interval and $\Delta t$ for the models trained with different dataset and tested on all datasets. **B**: The average MAPE and standard deviation (std) of MAPE are shown for the same models as in panel A.

We can see, that if the model is trained with data from the individual material groups, we get mostly good performance when testing within the same group (diagonal). Interestingly training on any of the nanocrystal groups will allow some learning transfer to the other nanocrystal groups. Indeed, if the model is trained on a dataset that includes different chemistries but keeps the same microstructures (NCs and TFs), we can see that the resulting models generally perform better, within the same microstructure-group. This learning transfer improves further, if we train on datasets separated by chemistries rather than microstructure (inorg. and hybr.). Then, the models perform well across all subgroups during testing. This is only improved further, if all data is used to train. To understand this behaviour, we also look at the std. of MAPE, which is an indicator of the similarity of train/test sets across splits, as aforementioned. In **Figure 5B** we plot av. and std. of MAPE together. There is a clear, linear correlation between the two, meaning that if the 30 train/test splits are more similar, then the performance during testing is also improved overall. In other words, if the train and test datasets are similar, the performance during testing is better. Again, we can use the statistical moments of the train set and test set can be used for this. In Figure S15 we show the absolute difference in mean, variance, skew, and kurtosis of the 81 trained with/tested with combinations. We find a strong indication that a lower difference in all moments leads to better performance, highlighting once again that the model is able to predict the behaviour of new, unseen samples well, if their behaviour is represented in the dataset during training. For instance, learning transfer between the NC groups is possible, because their degradation behaviour is comparable. Overall, we show that our approach can be used to learn more about the data structure, as well as the underlying material parameters.

## IV. CONCLUSION

In summary, we have established a photoluminescence (PL) stability forecasting algorithm based on a physics-informed segmentation approach. We demonstrated this method using a diverse data set composed of metal halide perovskites of different groups: inorganic and hybrid nanocrystals as well as thin films. A clear advantage of the physics-based approach is the explainability of the resulting model. The application of SHAP analysis helps to identify global and common descriptors for PL stability within the feature set. As a result, a direct link back to the underlying physical mechanisms governing material stability is enabled, specifically through the spectral shape and the relative behaviour of high- and low-energy regions, which helps to establish trust in the resulting predictions. This highlights the potential of using deep neural networks not only for performance prediction but also to enable new scientific insights from high-dimensional and complex data sets.

While the current results are promising, we anticipate further improvements of model performance through the extension of training data to include a wider range of chemistries and crystallinities. Equally, cases where the model struggles to generalise are themselves informative, as they may signal the presence of degradation phenomena not yet captured by the training data and thus warrant targeted investigation. The open code and training data for this work enables the broad adaptation within the scientific community.

## ACKNOWLEDGEMENTS

The authors acknowledge funding from the Swiss National Science Foundation (SNSF) (Project No. 226588), as well as funding from the Strategic Focus Area—Advanced Manufacturing (SFA-AM) through the project "Advanced Manufacturability of Hybrid Organic-Inorganic Semiconductors for Large Area Optoelectronics (AMYS)".

## CODE AVAILABILITY

The python code used in this work is readily available in our github repository[37].

# Supporting Information to:

# Physics-informed time-series forecasting of perovskite photoluminescence stability

Alexander Wieczorek[1,#], Manuel Kober-Czerny[1,#,*], Fábio Lopes[1,#], Austin George Kuba[2], Leon Müller[3], Christian M. Wolff[2], Jason Hattrick-Simpers[4,5], Sebastian Siol[1,*]

[1] Laboratory for Surface Science and Coating Technologies, Empa–Swiss Federal Laboratories for Materials Science and Technology, Ueberlandstrasse 129, Dübendorf CH-8600, Switzerland.

[2] Institute of Electrical and Microengineering (IEM), Photovoltaic and Thin-Film Electronics Laboratory, EPFL –École Polytechnique Fédérale de Lausanne, Switzerland.

[3] Fritz-Haber-Institut der Max-Planck-Gesellschaft, Faradayweg 4-6, D-14195 Berlin, Germany

[4] Department of Materials Science and Engineering, University of Toronto, M5S 3E4, Toronto, ON, Canada.

[5] Acceleration Consortium, University of Toronto, M7A 2S4, Toronto, ON, Canada.

[s#] These Authors contributed equally

Corresponding Authors: Manuel.Kober-Czerny@empa.ch, Sebastian.Siol@empa.ch

## Materials & Methods

### Deposition of perovskite libraries

All perovskite libraries were deposited on 1.1 mm thick borosilicate glass (EXG) substrates sized 5×5 $cm^2$. Initial cleaning of the substrates was performed through ultrasonic cleaning in acetone, and ethanol, with subsequent drying in a $N_2$ flow.

### $FAPbBr_3$ nanocrystal deposition libraries

Gradients of $FAPbBr_3$ nanocrystals were obtained through gradient deposition of commercially available perovskite quantum dot dispersions (Avantama Q-520): 500 µl dispersion volume was first rapidly drop-casted onto the EXG substrate. Then, drying of the film was performed using a $N_2$ flow gun. By angling the heat gun during the drying step, a pronounced thickness gradient was obtained, as verified through optical absorption measurements of the resulting library.

### High entropy alloyed nanocrystal composition libraries

Compositionally diverse high entropy alloyed nanocrystal libraries were obtained through post-synthetic treatment of $CsPbBr_3$ nanocrystals, as reported elsewhere[4]. Likewise, the used samples and degradation data was appropriated from the seminal publication.

### FA-Cs-Pb-Br-I compositional libraries

The hybrid deposition process was adapted from previous work[1], [2], [3].

A Lesker Mini SPECTROS system was first used to co-evaporate the inorganic components of the perovskite. The deposition rates were monitored by quartz crystal monitors (QCMs), which were initially calibrated by individually evaporating CsBr and $PbI_2$ using a rotating substrate holder, checking the thickness using cross-section SEM and adjusting the tooling factor appropriately. For the deposition of the films in the study, the substrate was held stationary and the substrate was placed in the center of the deposition plate, approximately equidistant from each source. This naturally leads to a gradient in the $CsBr:PbI_2$ ratio and thickness across the piece that can be exploited to form a library of different compositions. Shadow masks with a 5×9 grid were used to create 45 well-defined and spatially separate pixels per substrate. One pixel was used for reference measurements, leaving 44 perovskite samples that could be tracked over aging time. The nominal deposition rates as determined by the QCMs were 0.1 Å $s^{-1}$ CsBr and 1.0 Å $s^{-1}$ $PbI_2$ for a nominal thickness of 198 nm.

To convert the inorganic templates into hybrid perovskites, a solution of 2.9:1 mol FAI:FABr in ethanol (50 mg $ml^{-1}$ total) was dynamically spin coated onto the inorganic template films at 4000 rpm for 30 s in a nitrogen atmosphere. The films were dried on a hotplate at 80°C in a $N_2$ atmosphere for 5 minutes. Next, the films were removed to ambient atmosphere and then annealed at 150 °C for 30 minutes in controlled humidity box with 15% relative humidity. Finally, the films were rinsed with IPA (dynamic spin coating at 4000 rpm for 30 s) without further post drying to remove residual organohalides.

***Cs-Pb-Br-I compositional libraries***

In addition to the cleaning steps described earlier, substrates were first ultrasonically cleaned in a 3% solution of HELLMANEX III (Hellma Analytics) in deionized (DI) water for 10 min.

The substrates were masked and mounted in the evaporation chamber. Masks were used to generate a pattern with 5×9 pixels. With one reference pixel per substrate covered, this resulted in 44 perovskite thin films with different compositions per substrate.

The co-evaporation was performed in a Lesker Mini SPECTROS, integrated into a glovebox under an inert N2 atmosphere (<1 ppm $O_2$ and $H_2O$). The precursor salts were cesium bromide (CsBr, 99.9 % metal basis, abcr), lead iodide ($PbI_2$, 99.999 % trace metal basis, Sigma Aldrich), and cesium iodide (CsI, 99.999 % metal basis, ThermoScientific).

To reach the compositional gradients, the substrate holder was kept stationary. Before the evaporation was started, the evaporation chamber was pumped down until a pressure of less than $2\cdot10^{-6}$ mbar was reached. The precursors were preheated, and a stable rate (two digits precision) was acquired before the substrate shutter was opened. The rates were chosen to be 0.15, 0.49 and 0.3 Å $s^{-1}$ for CsBr, PbI2 and CsI, respectively, corresponding roughly to a molar ratio of 1:2:2 in the rotating samples. Each evaporation source is equipped with a quartz crystal microbalance.

The deposition was started upon reaching stable rates for all sources and stopped, when reaching a nominal thickness of 180 nm of $PbI_2$ (tooled for rotating substrates at a fixed distance).

**Table S1** Overview of the datasets of the different metal halide perovskite groups used in this work.

| Group | inorg. NC | inorg. TF | hybr. NC | hybr. TF |
|---|---|---|---|---|
| composition | High-entropy alloyed $CsPbBr_3$ derivatives[4] | $Cs_aPb_bBr_cI_d$ | $FAPbBr_3$ with deposition gradients | $FA_aCs_bPb_cBr_dI_e$ |
| crystallinity | nanocrystalline | polycrystalline | nanocrystalline | polycrystalline |
| ageing temperature | 25 °C | 85 °C | 60 °C | 85 °C |
| ageing atmosphere | ambient | $N_2$ | $N_2$ | $N_2$ |
| **raw data** | | | | |
| samples | 34 | 44 | 45 | 44 |
| timepoints | 5316 | 11616 | 34785 | 34364 |
| **used data** | | | | |
| samples | 32 | 42 | 45 | 44 |
| timepoints | 3812 | 5894 | 6750 | 4000 |
| **for CNN-LSTM** | | | | |
| training set | 20 samples | 26 samples | 28 samples | 28 samples |
| validation set | 5 samples | 7 samples | 8 samples | 7 samples |
| test set | 7 samples | 9 samples | 9 samples | 9 samples |

**Fitting the PL spectra**

We processed the time-series photoluminescence (PL) spectra as obtained from the spectrometer for each material group individually. First, since every experiment starts at room-temperature and there is some time, we removed all time-points before the desired ageing temperature (see Table S1) was reached. The x-values were then transformed from wavelength into energy (eV) and the y-values were changed using the Jacobian transformation. The spectra were afterwards cropped into a window surrounding the PL peak (1.4-2.4 eV for the thin film samples; 2.0-2.8 eV for the nanocrystal samples). From each spectrum, we subtracted a linear background by fitting a straight line between the edges (10 data points from each edge) of the cropped spectra.

To estimate the noise of our data ($\sigma_{\text{noise}}$), we applied a strong Savitzky-Golay filter (scipy.signal.savgol_filter(51,2)) and obtained $y_{\text{smooth}}$, such that

$$\sigma_{\text{noise}} = \sqrt{\frac{1}{N}\sum_{i=1}^{N}(y_{\text{smooth}} - y)^2} \tag{S1}$$

is the standard deviation.
Then, for each time-series, we checked for the existence of a main PL peak using the find_peaks function from the scipy.signal package in python on $y_{\text{smooth}}$. To count as a peak, its prominence had to be at least $2\sigma_{\text{noise}}$ and its width at least 50 datapoints. If no peak was found, the algorithm returned NaNs.
If a peak was found, then its position was the initial guess for high-energy (HE) peak center. The unsmoothed data was then fitted to a normalized, pseudo-Voigt model

$$f(x, A, \mu, \sigma, \alpha) = \mathrm{A} \cdot \left(\frac{(1-\alpha)}{\sigma_g\sqrt{2\pi}} \cdot e^{\left(-\frac{(x-\mu)^2}{2\sigma_g{}^2}\right)} + \frac{\alpha}{\pi} \cdot \frac{\sigma}{(x-\mu)^2 + \sigma^2}\right), \tag{S2}$$

where $\mu$ is the center of the HE peak (HE peak center), $A$ is its area (HE peak area), $\sigma$ is the full-width at half-maximum (HE peak fwhm), and $\sigma_g = \sigma/\sqrt{2\ln 2}$ is the broadening of the gaussian. $\alpha$ is the fraction of the lorentzian component of the total peak. During the fitting, the residuals are weighted by

$$w_i = \begin{cases} e^{|x_i - c|}, & x \geq c \\ {y_i}/{\max{(y)}}, & x < c' \end{cases} \tag{S3}$$

where $c$ is HE peak center as obtained from find_peaks. This was necessary to consistently make the fitted peak align with the high-energy side of the PL peak.
Afterwards, the reduced $\chi^2$ is calculated as a fit metric (HE peak $\chi^2$) and the signal-to-noise ratio (S to N) is estimated as

$$\frac{S}{N} = \frac{\max{(f(x, A, \mu, \sigma, \alpha))}}{\sigma_{\text{noise}}}. \tag{S4}$$

Then, the residual area of the spectrum and the HE peak is then ascribed to a low-energy (LE) feature (LE tail area). Here, we only integrated over the unsmoothed data, where the y-values were larger than $\sigma_{\text{noise}}$. We then calculated the ratio of the LE and HE areas (LE to HE ratio). Lastly, the unsmoothed, cropped PL spectrum was integrated to obtain the total area, which was later normalized to value at $t = 0$ to obtain Norm area. We summarize all extracted features in Table S2.

**Table S2** Summary of the extracted features and an explanation of each.

| No. | Feature | Explanation |
|---|---|---|
| | **HE peak** | The PL data was fitted with a normalized pseudo-Voigt model, which combines a gaussian and lorentzian peak, both with the same fwhm. |
| 1 | HE peak center | center of the HE peak (in eV) |
| 2 | HE peak fwhm | fwhm of the gaussian and lorentzian part |
| 3 | HE peak area | area of the HE peak |
| 4 | HE peak $\chi^2$ | reduced $\chi^2$ metric after the HE peak fit |
| | **LE feature** | Afterwards, the residual area after the pseudo-Voigt fit is ascribed to a low-energy PL feature. |
| 5 | LE tail area | area low energy feature |
| | **Other** | |
| 6 | LE to HE ratio | area ratio of LE tail area/HE peak area |
| 7 | S to N | signal-to-noise ratio calculated as the ratio of the maximum of the HE PL peak and the estimated noise ($\sigma_{\text{noise}}$) |
| 8 | Norm_area[a.] | Total_area normalized to the value at *t* = 0 for each timeseries |

**Assessing the Performance of the Model**

To compare the performance of the CNN-LSTM model for different datasets we use a few metrics in the manuscript and supporting information. The first one, which is used to assess the model performance during the training and validation is the mean square error (MSE) and is defined as

$$\mathrm{MSE} := \frac{1}{n} \cdot \sum_{i=1}^{n} (M_i - P_i)^2, \tag{S5}$$

where $M_i$ and $P_i$ are the measured and predicted values, respectively, and $n$ is the total number of datapoints. To assess the performance of the models during testing, we use several metrics, such as the root mean square error (RMSE)

$$\mathrm{RMSE} := \sqrt{\mathrm{MSE}} = \sqrt{\frac{1}{n} \cdot \sum_{i=1}^{n} (M_i - P_i)^2}, \tag{S6}$$

the mean absolute percentage error (MAPE)

$$\mathrm{MAPE} := \frac{1}{n} \cdot \sum_{i=1}^{n} \left| \frac{M_i - P_i}{M_i} \right|, \tag{S7}$$

and the coefficient of determination ($R^2$)

$$\mathrm{R}^2 := 1 - \frac{\sum_{i=1}^{n} (M_i - P_i)^2}{\sum_{i=1}^{n} (M_i - \bar{M})^2}, \tag{S8}$$

where $\bar{M}$ is the mean of the measured data.

### Statistics of the Datasets

Here, we look at some statistical moments of the different datasets used for training to try to understand why the model trained on some data will perform well and when trained on others not. For this, we compare the average and standard deviation (std) of the MAPE of the 30 train/test/validation splits for each dataset used in **Table S3**. Then, we use the Norm area feature, which is the value to be predicted from the model and calculate the mean, variance, skew, and kurtosis (normal distribution at 0).
We can see that a lower average of MAPE generally coincides with a lower std of MAPE, as well as a lower variance and skew of the dataset. In Figure S14 we show how the different moments are correlated with the av. and std. of MAPE.

**Table S3** Statistics of the dataset as demonstrated by the Norm area feature. The mean absolute percentage error (MAPE) of the CNN-LSTM model trained on the individual subsets, as well as all data is shown. The standard deviation (std) is calculated across the 30 train/test/validation splits.

| Dataset | after training CNN-LSTM | | statistics of Norm_area feature data | | | |
|---|---|---|---|---|---|---|
| | av. of MAPE | std. of MAPE | mean | variance | skew | kurtosis |
| **inorg. NC** | 0.078 | 0.032 | 0.732 | 0.233 | 0.017 | -1.075 |
| **inorg. TF** | 0.221 | 0.040 | 0.461 | 0.811 | 1.052 | -0.181 |
| **hybr. NC** | 0.098 | 0.033 | 0.787 | 0.360 | 0.287 | -0.303 |
| **hybr. TF** | 0.147 | 0.021 | 0.502 | 0.774 | 1.774 | 2.170 |
| **NCs** | 0.084 | 0.024 | 0.767 | 0.326 | 0.394 | 0.131 |
| **TFs** | 0.209 | 0.034 | 0.478 | 0.797 | 1.367 | 0.955 |
| **inorg.** | 0.174 | 0.033 | 0.568 | 0.593 | 0.335 | -0.844 |
| **hybr.** | 0.110 | 0.025 | 0.681 | 0.521 | 0.614 | -0.247 |
| **all data** | 0.138 | 0.019 | 0.628 | 0.559 | 0.497 | -0.363 |

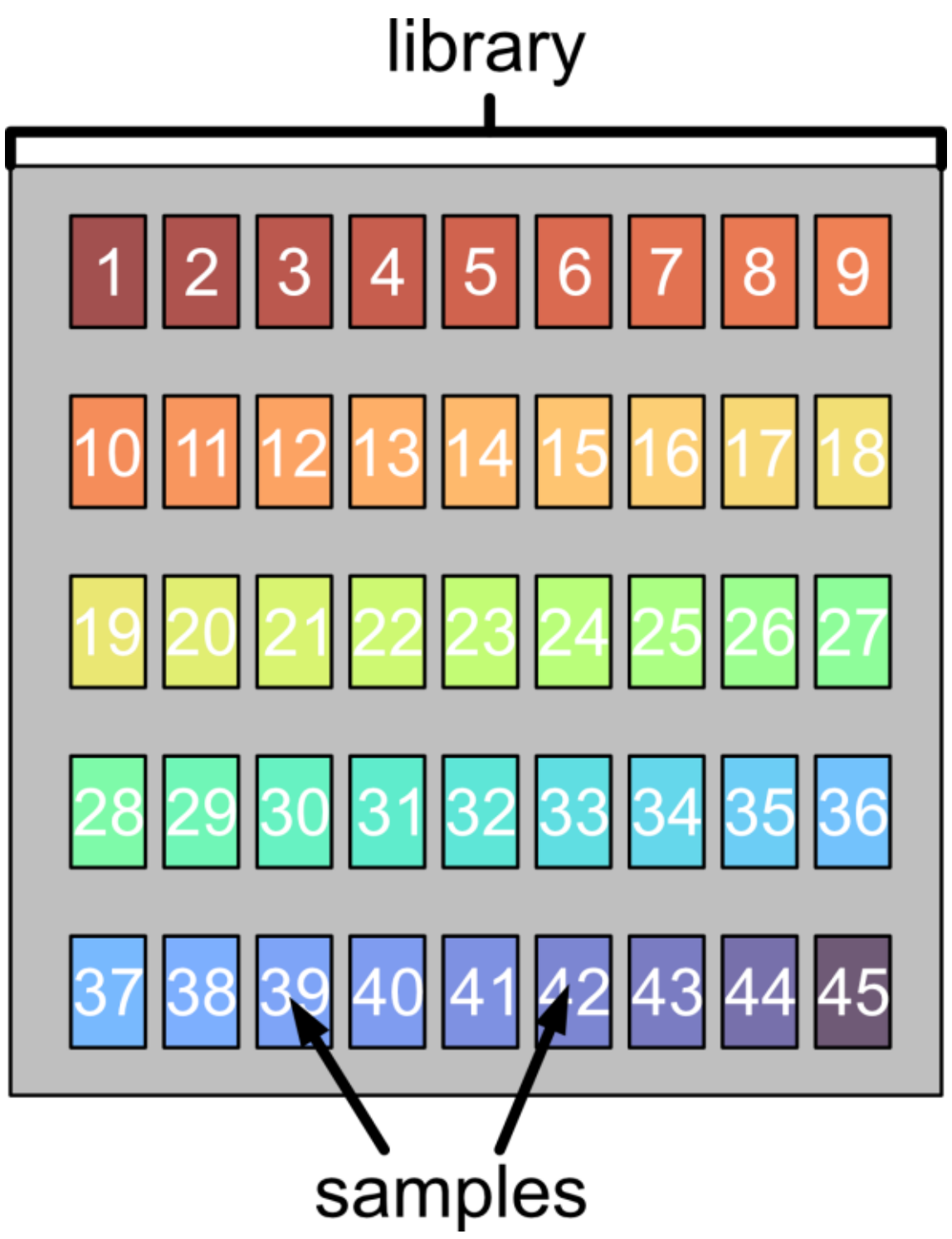

**Figure S1** A typical layout of a library of a material group with the different samples on it as used in this work. The colors match the ones in Figure 2 in the main text, as well as Figures S4-S7.

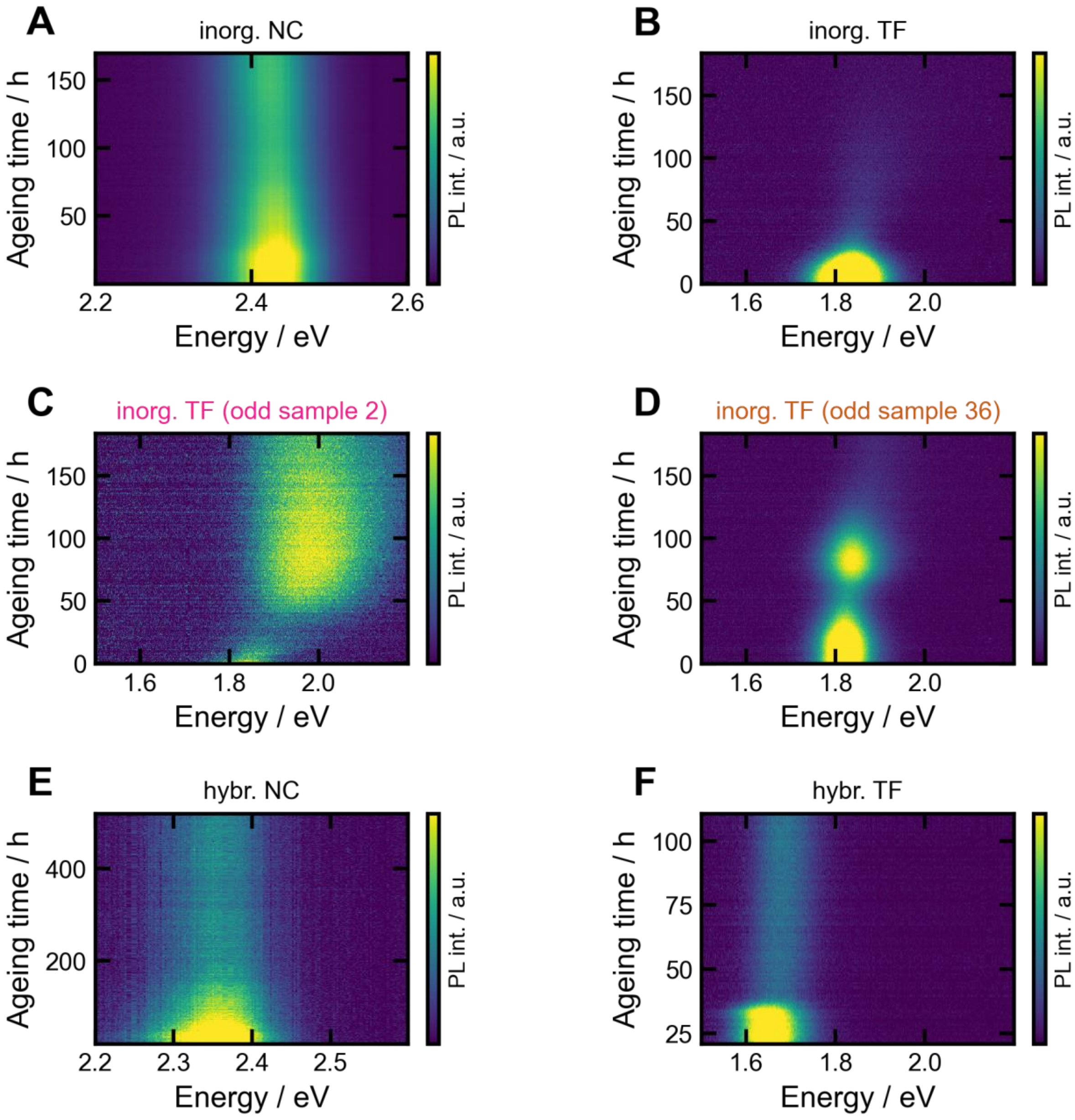

**Figure S2** Time-dependent PL spectra of inorg. NC(**A**), inorg. TF (**B**,**C**,**D**), hybr. NC (**E**), and hybr. TF (**F**). For each material the representative sample number 23 (the center of the library) has been chosen here. In case of the inorg. TF, sample number 2 (upper left corner) and sample number 36 (towards the lower right) are shown as well as representative 'extreme' compositions. Their positions on the inorg. TF library can be seen in Figure S2.

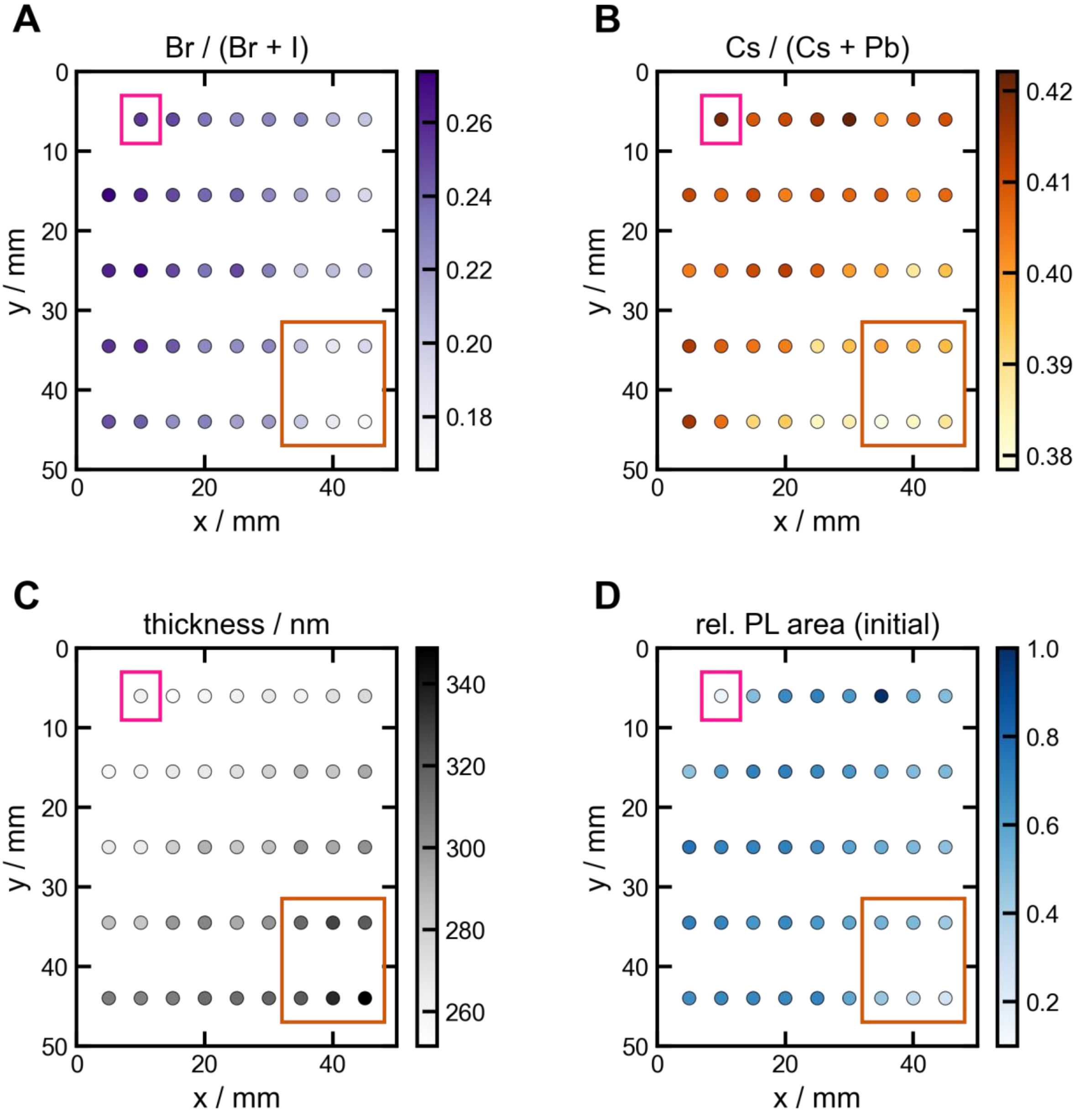

**Figure S3 A**: Bromide fraction, **B**: Cesium fraction, and **C**: thickness of the different samples of the inorg. TF library are shown as obtained from XRF. **D**: The PL area at $t$ = 0 h is shown (normalized to the highest value of the library) for the same samples on the library. The pink box marks sample number 2, while the orange box marks samples number 34-36 and 43-45. Both groups of samples show different behaviour to the group mean in the normalized PL area as shown in Figure 2 in the main text.

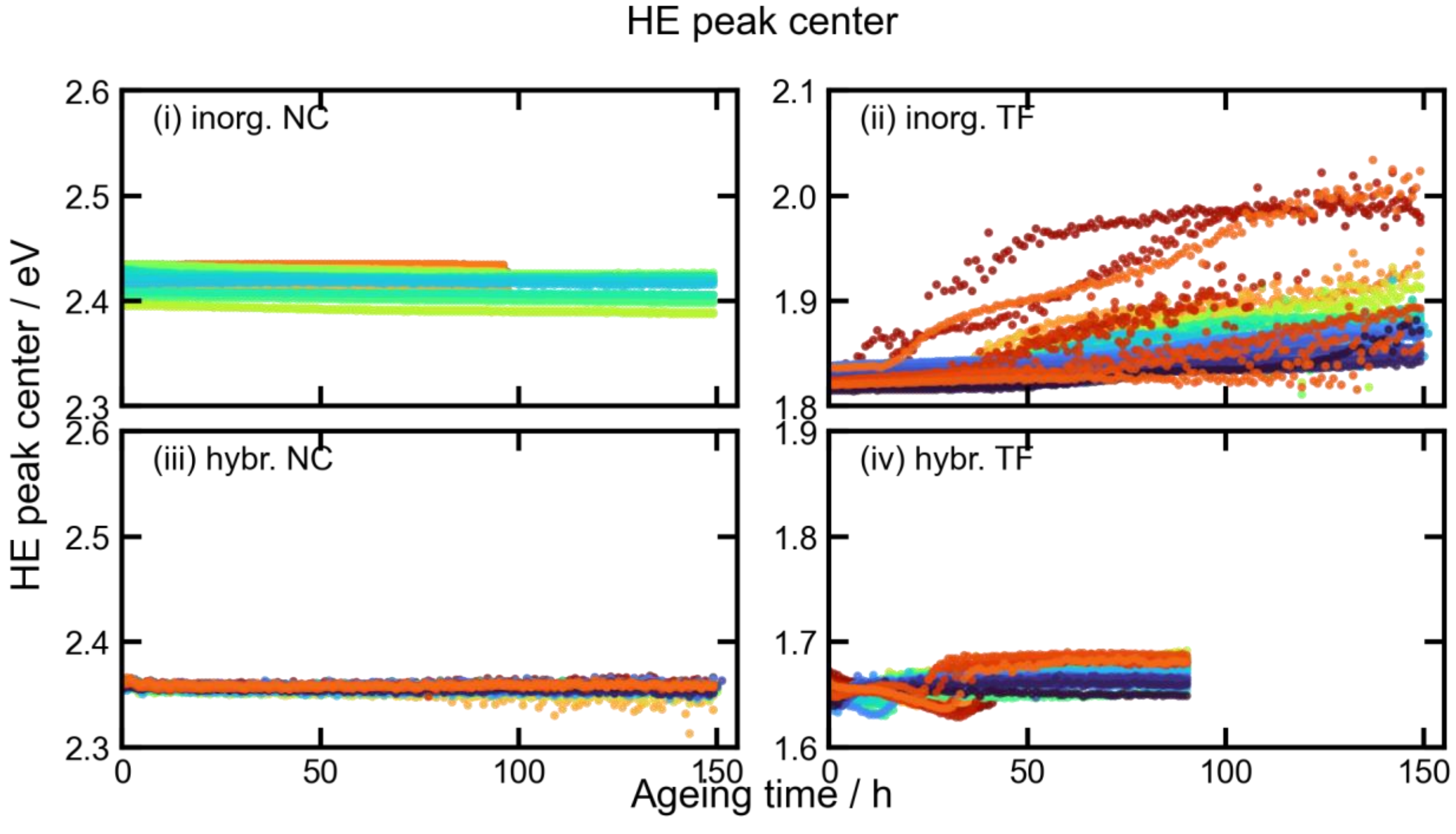

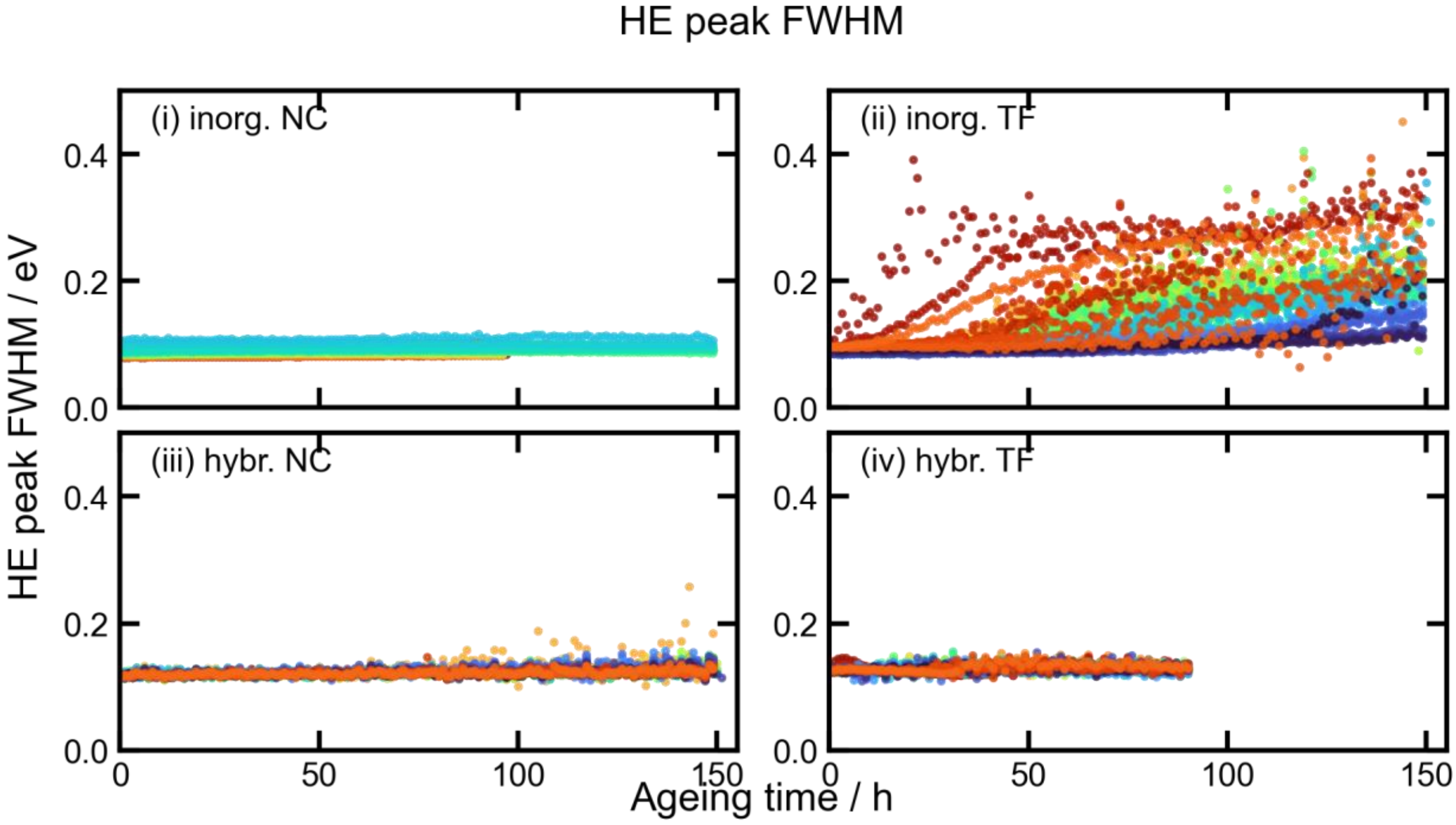

**Figure S4** The HE peak center (**top**) and HE peak fwhm (**bottom**) are shown for the 4 different groups of samples used in this study ((i)-(iv) as a function of ageing time.

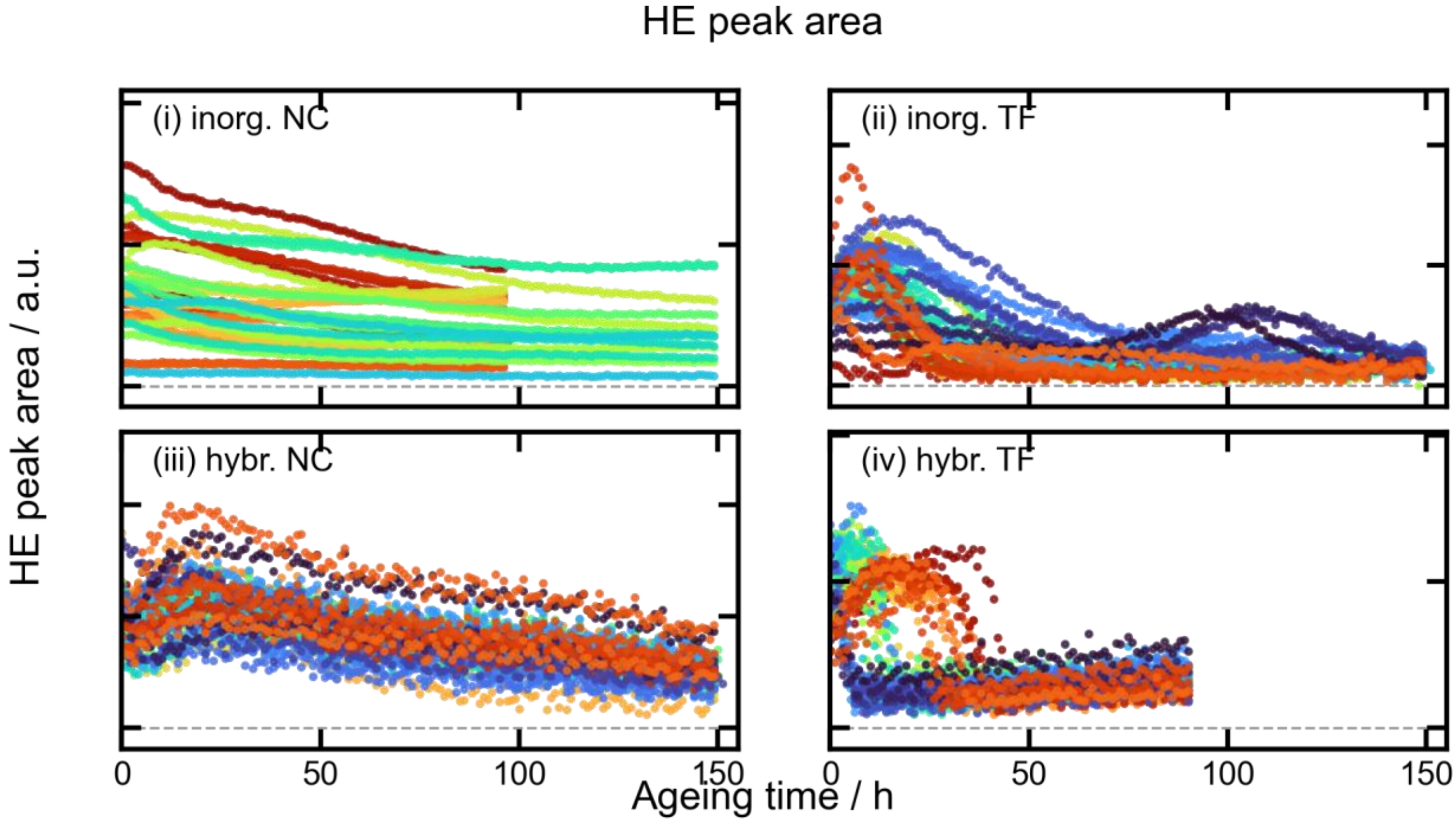

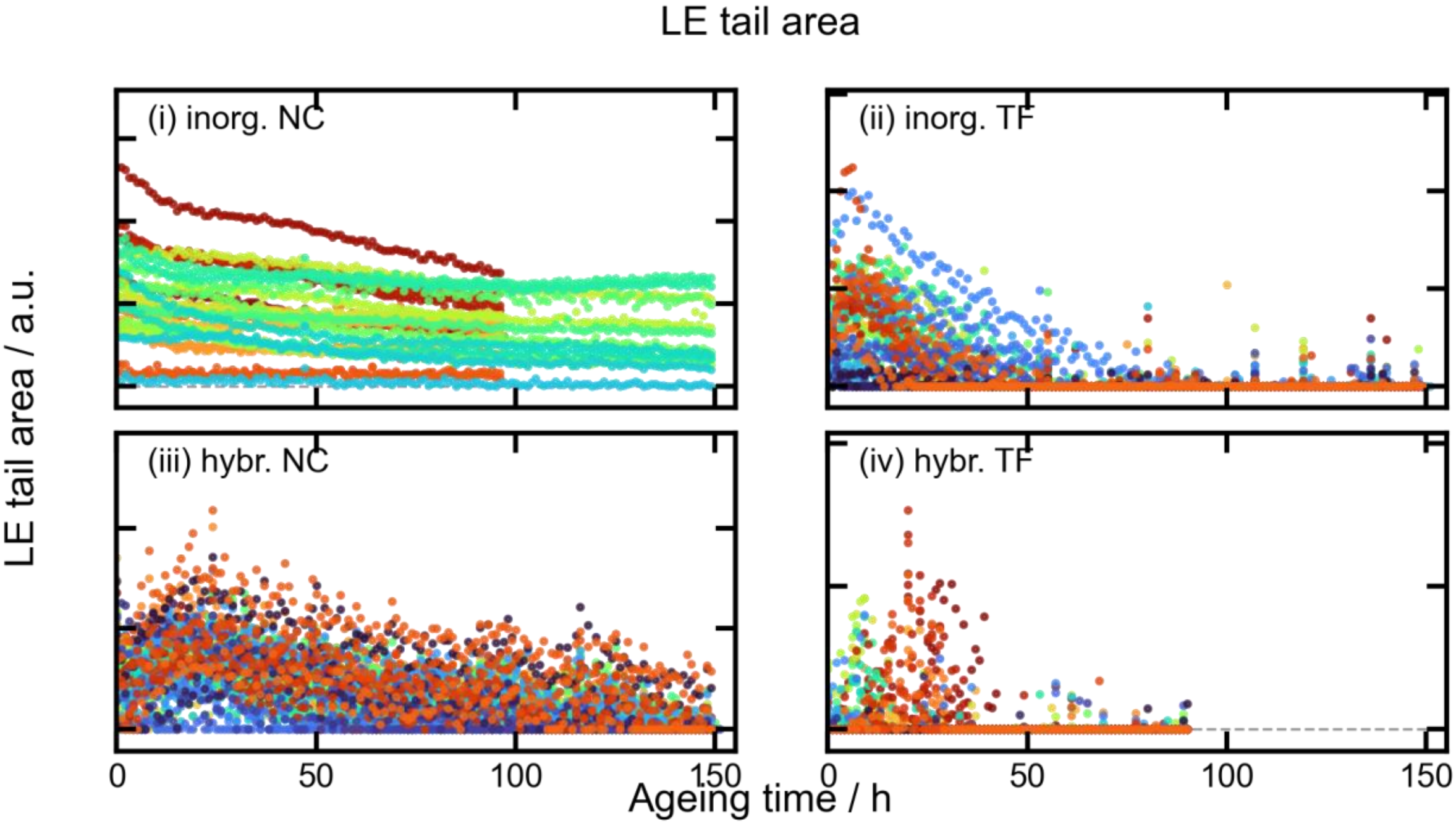

**Figure S5** The HE peak area (**top**) and LE tail area (**bottom**) are shown for the 4 different groups of samples used in this study ((i)-(iv) as a function of ageing time. The dotted line indicates the zero level.

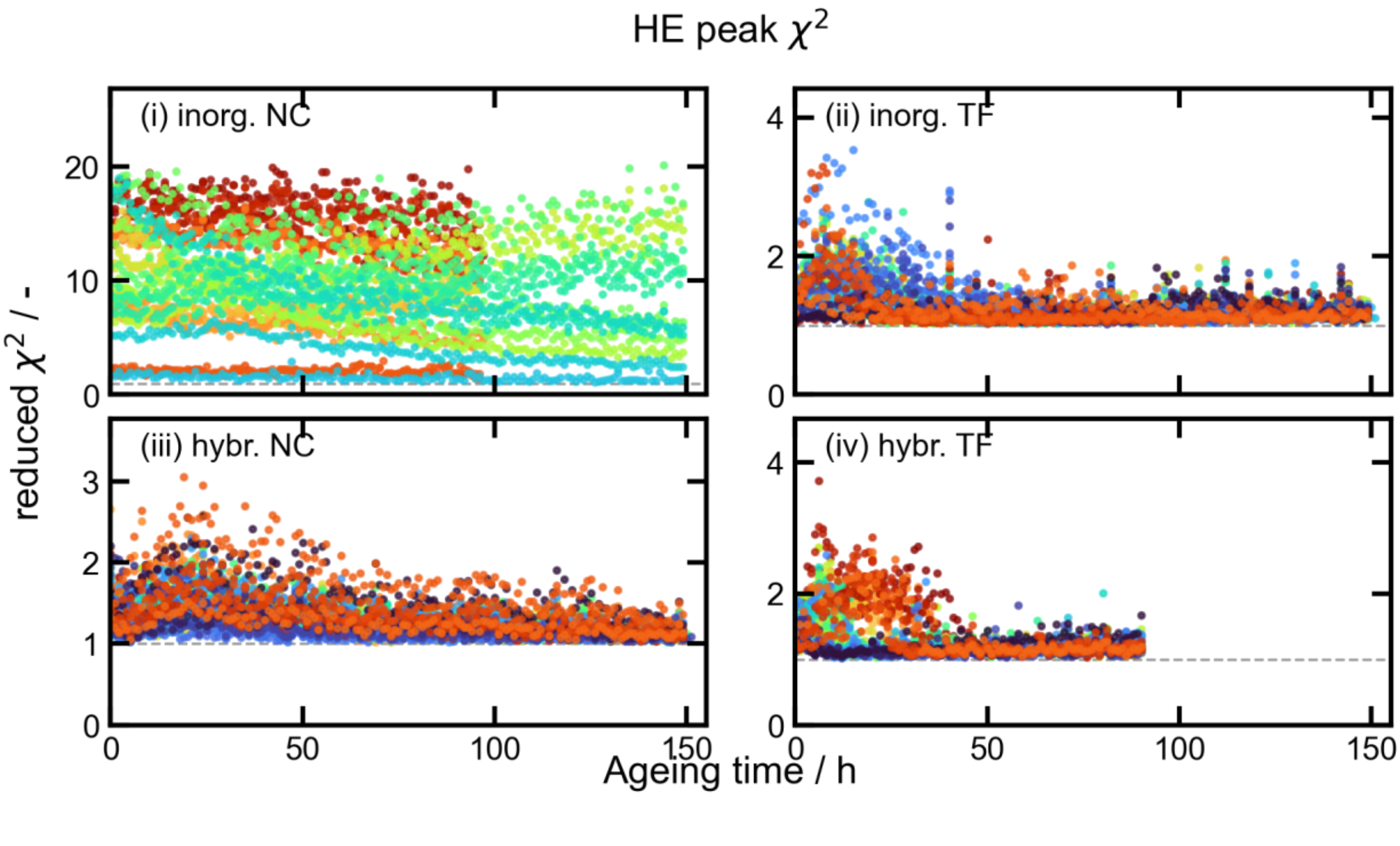

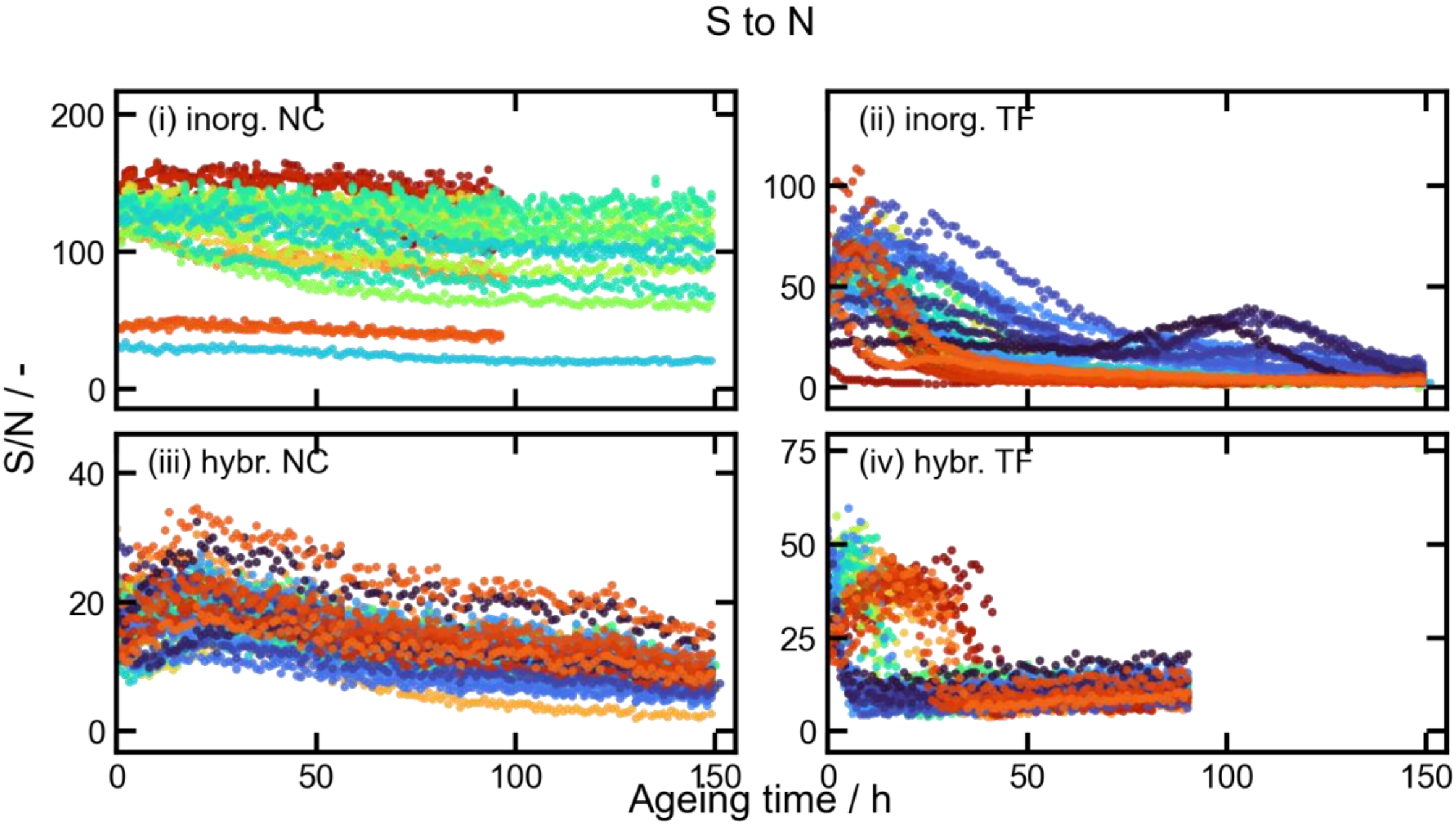

**Figure S6** The HE peak $\chi^2$ (**top**) and S to N ratio (**bottom**) are shown for the 4 different groups of samples used in this study ((i)-(iv) as a function of ageing time. The dotted line indicates the zero level.

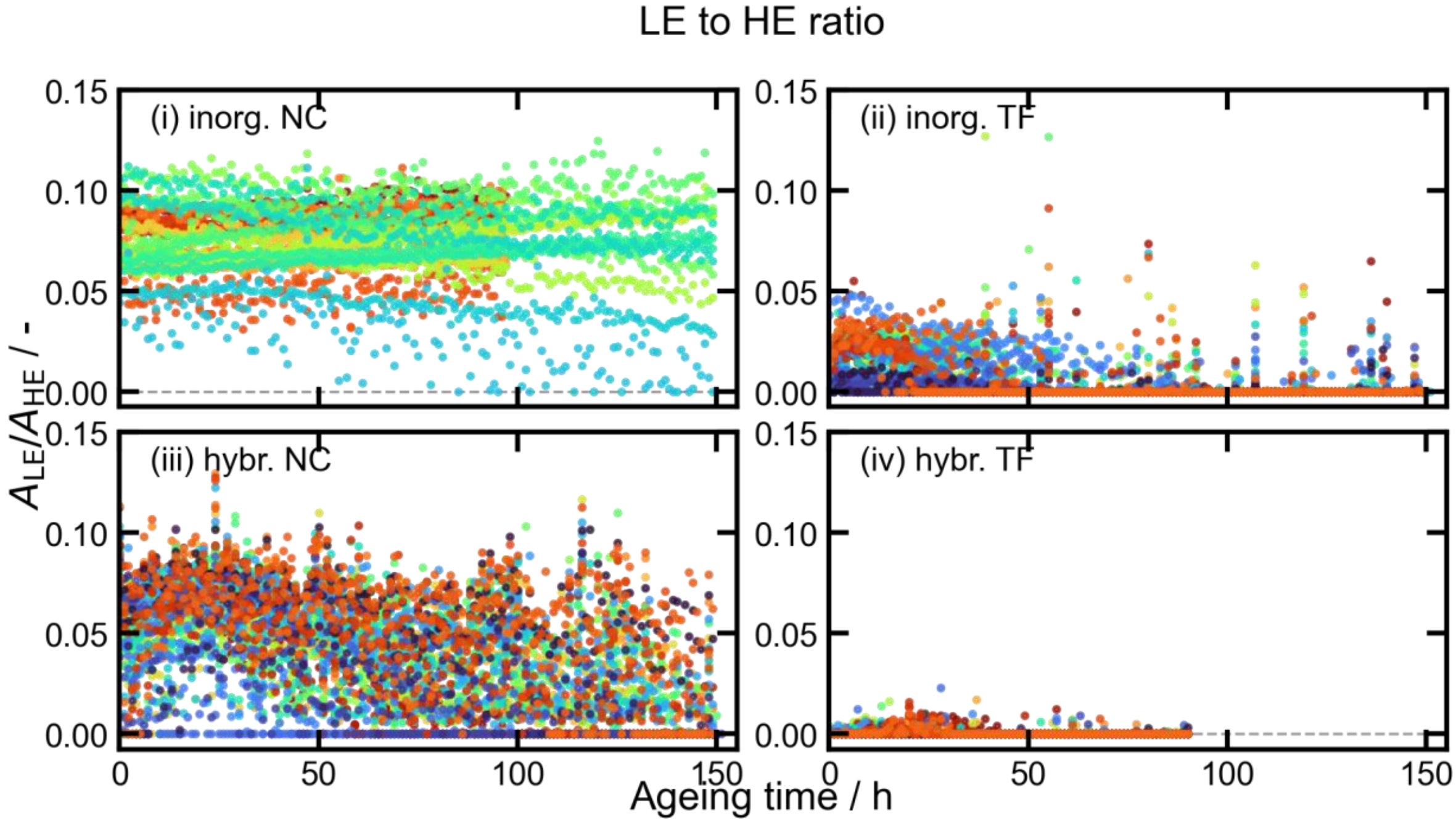

**Figure S7** The LE to HE ratio is shown for the 4 different groups of samples used in this study ((i)-(iv) as a function of ageing time. The dotted line indicates the zero level.

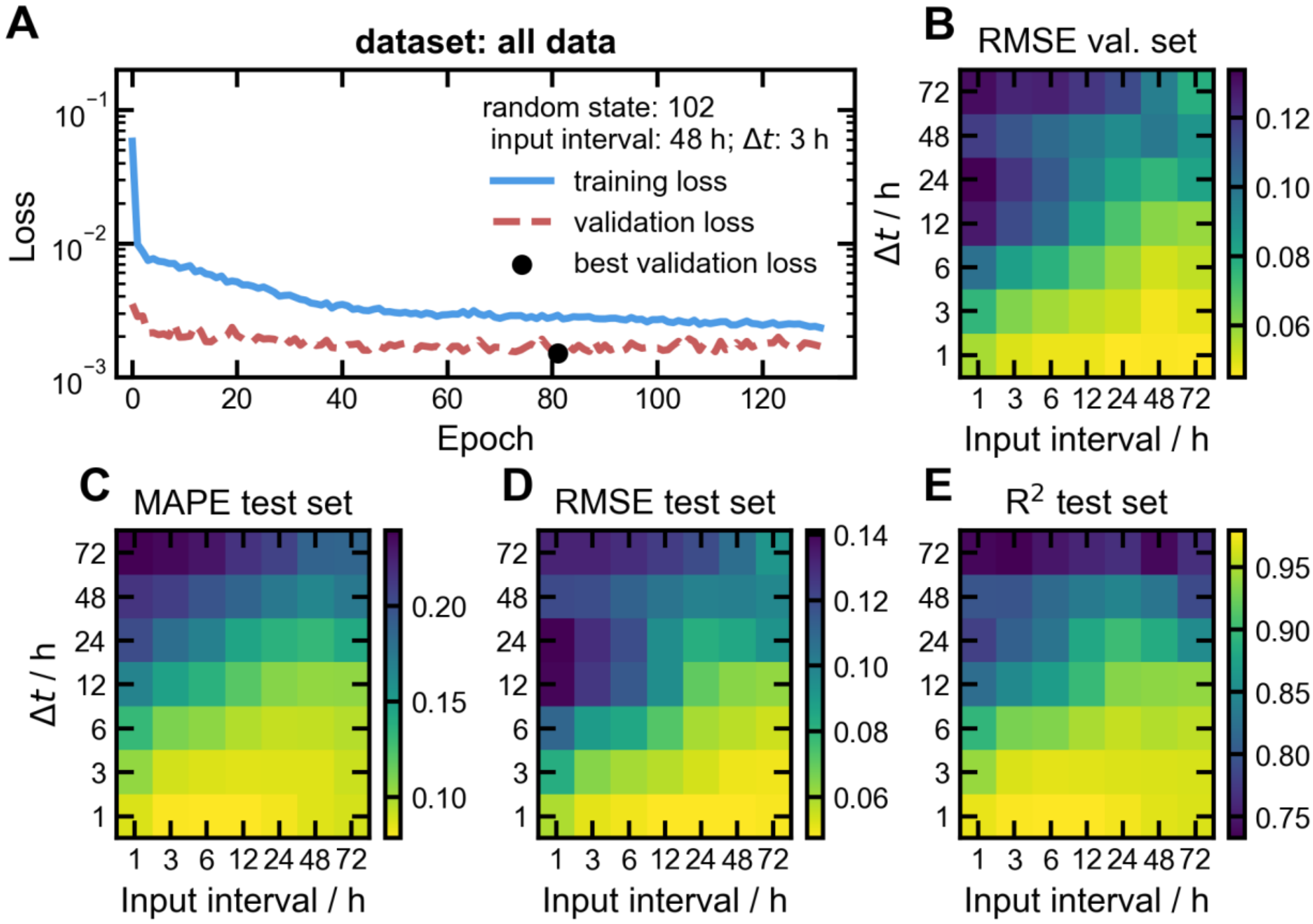

**Figure S8 A**: Train and validation loss curves for a representative model trained on all data (random state = 102 for the train/test/validation split). The curves are shown for the input interval/Δ*t* of 48/48 hours. **B**: The average root mean square error (RMSE) is shown for the models trained with all data (across all train/test/validation splits) for the whole 7x7 grid of input intervals and Δ*t*'s. **C-E**: The performance of the model is assessed using a test set of unseen data. Then the mean absolute percentage error (MAPE, **C**), the RMSE (**D**), and the coefficient of determination ($R^2$ score, **E**) are shown for the 6x6 grids. Again, only the average values of the 30 train/test/validation splits are shown.

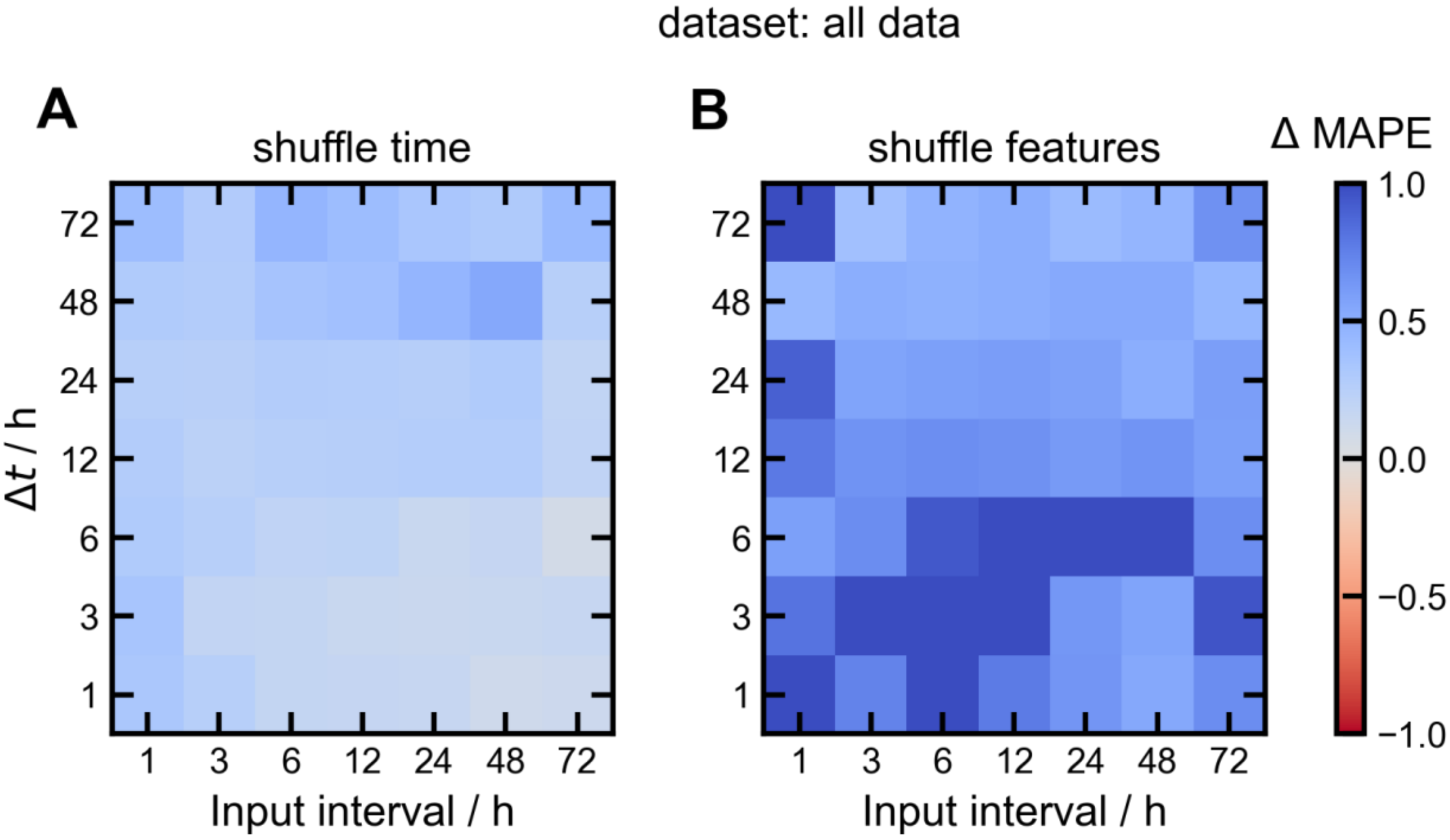

**Figure S9** The difference in mean absolute percentage error (MAPE) is shown as an average of the 30 train/test/validation splits for the different shuffling benchmarks used: **A**: the shuffled time axis before making the windows, **B**: the shuffling of the order of the features with the regular windows and time axis.

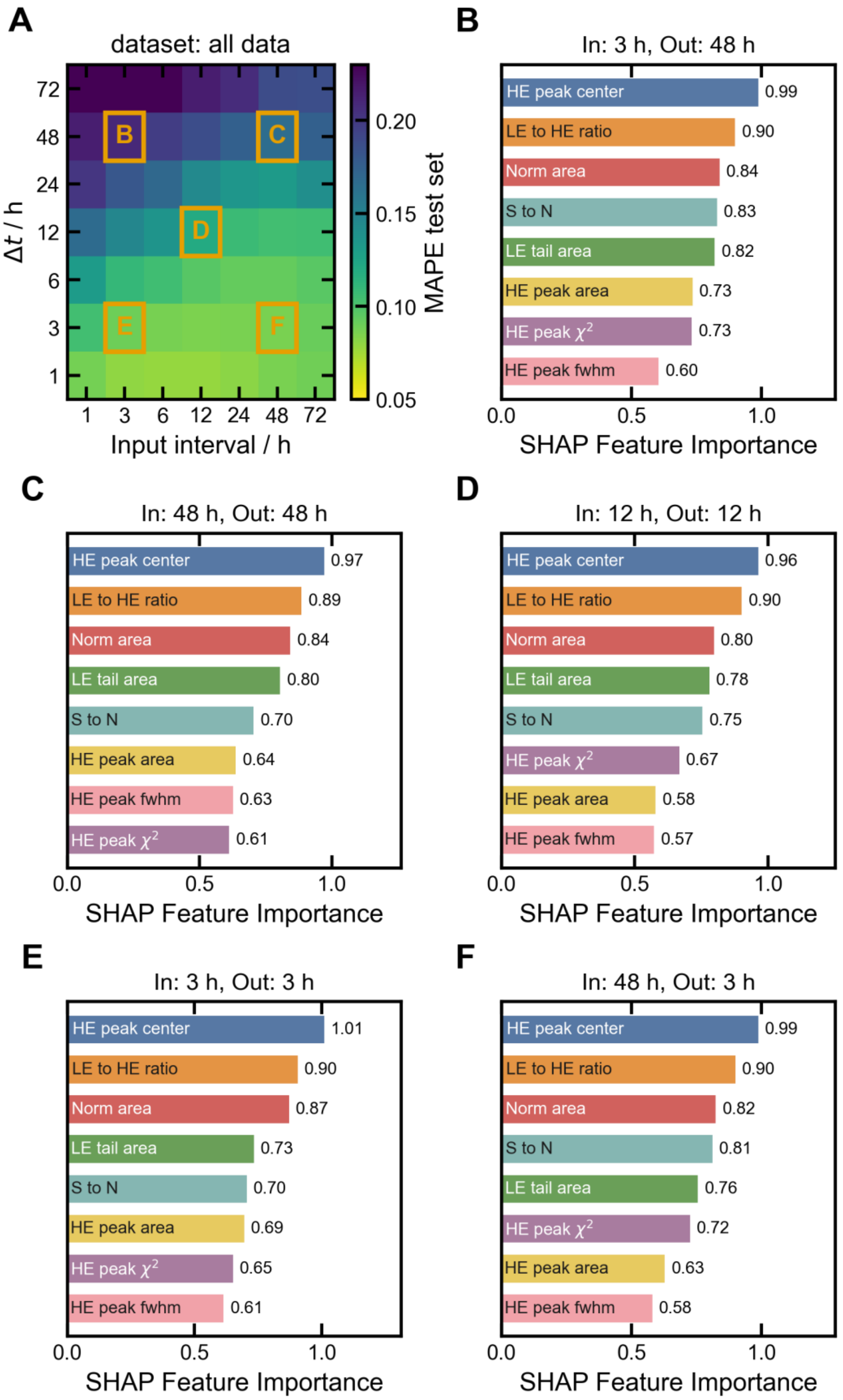

**Figure S10 A**: The mean absolute percentage error (MAPE) is shown as an average of the 30 train/test/validation splits. The orange boxes mark the input interval/$\Delta t$ combinations that are used for the SHAP analysis: **B**: 3/48, **C**: 48/48, **D**: 12/12, **E**: 3/3, and **F**: 48/3. The colours for the different features are kept consistent to allow better comparisons.

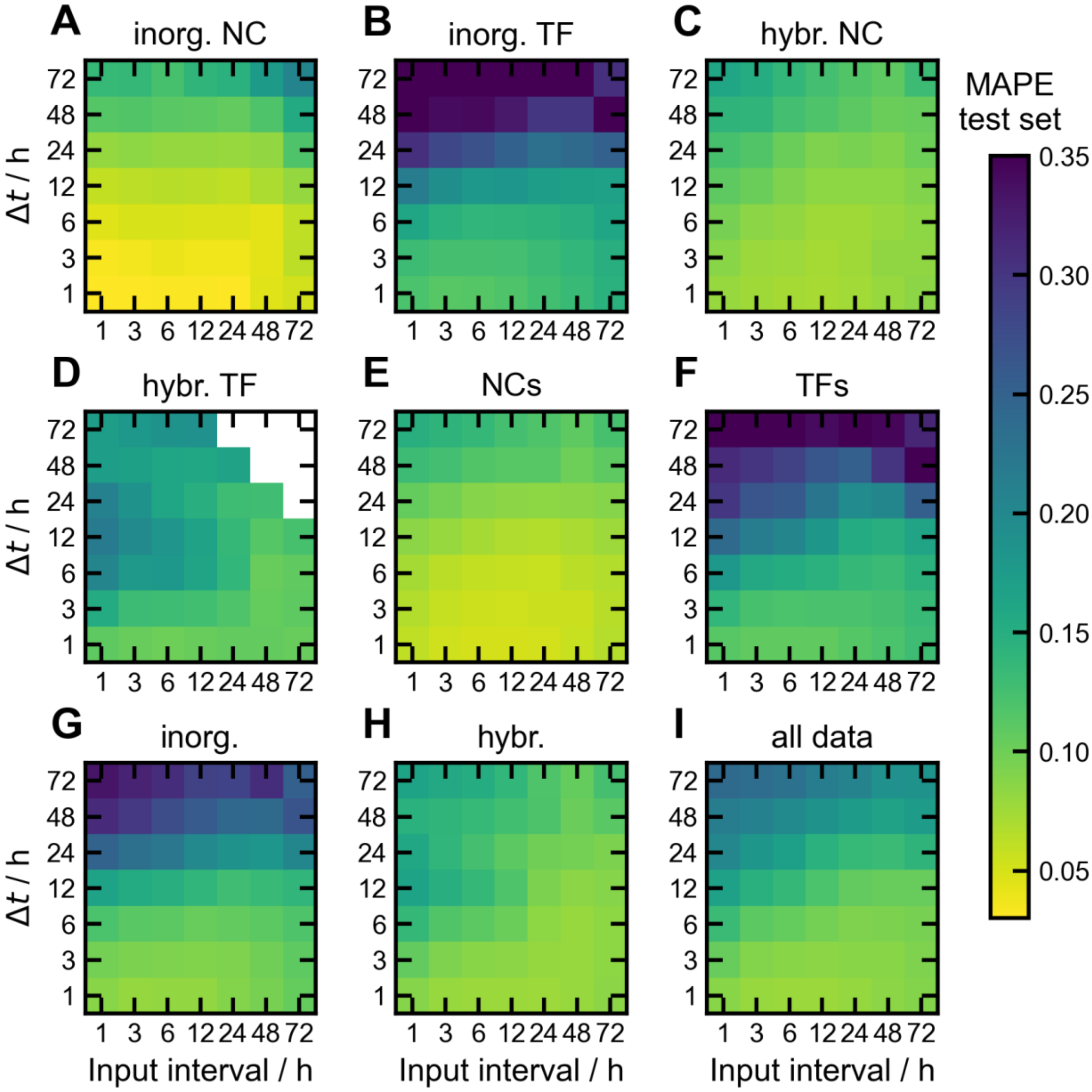

**Figure S11** The mean absolute percentage error (MAPE) is shown as an average of the 30 train/test/validation splits for the models trained on the individual subsets of the data: **A**: only the inorganic nanocrystals, **B**: only the inorganic thin films, **C**: only the hybrid nanocrystals, **D**: only the hybrid thin films, **E**: both nanocrystal sets, **F**: both thin film sets, **G**: both inorganic sets, **H**: both hybrid sets, and **I**: all the data combined.

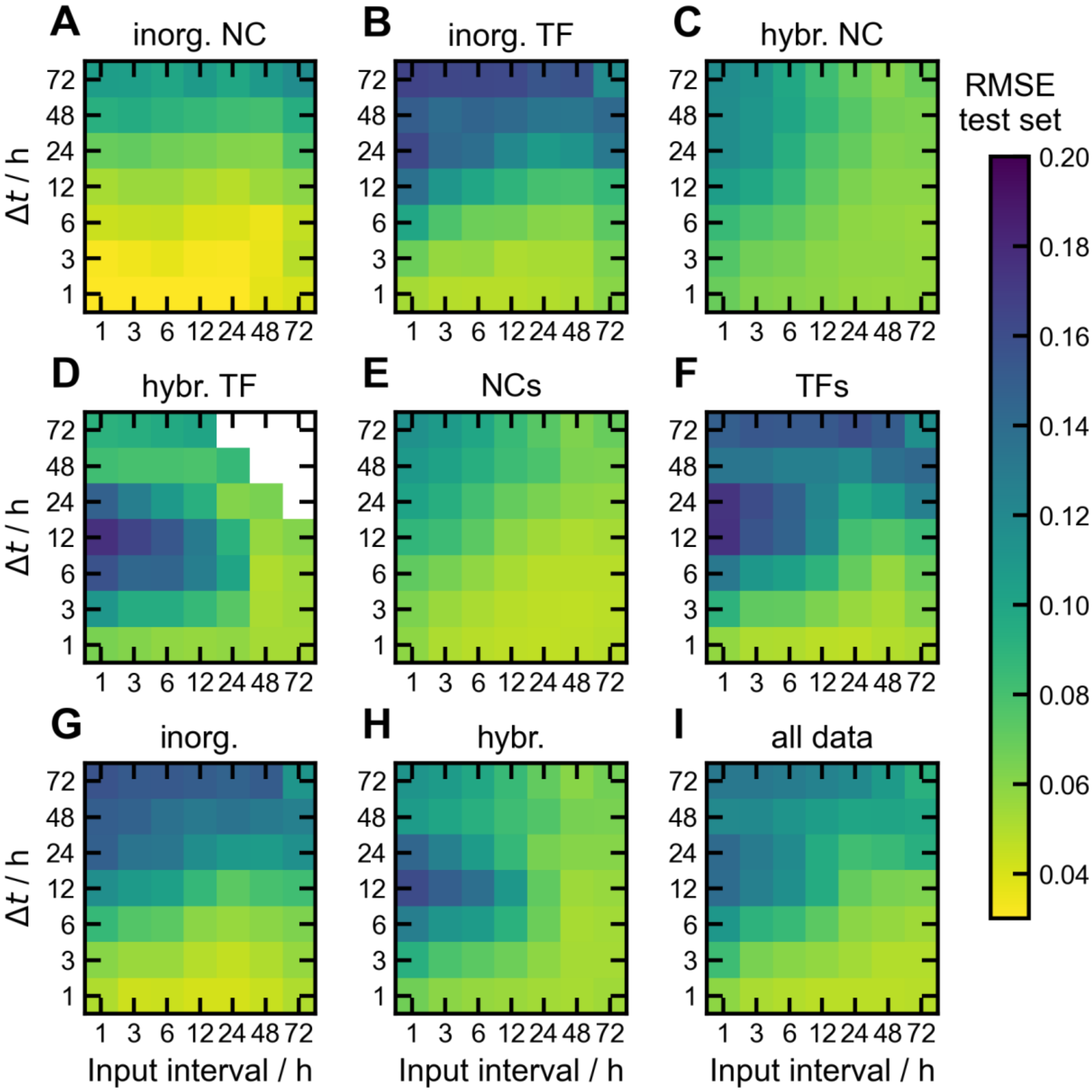

**Figure S12** The root mean square error (RMSE) is shown as an average of the 30 train/test/validation splits for the models trained on the individual subsets of the data: **A**: only the inorganic nanocrystals, **B**: only the inorganic thin films, **C**: only the hybrid nanocrystals, **D**: only the hybrid thin films, **E**: both nanocrystal sets, **F**: both thin film sets, **G**: both inorganic sets, **H**: both hybrid sets, and **I**: all the data combined.

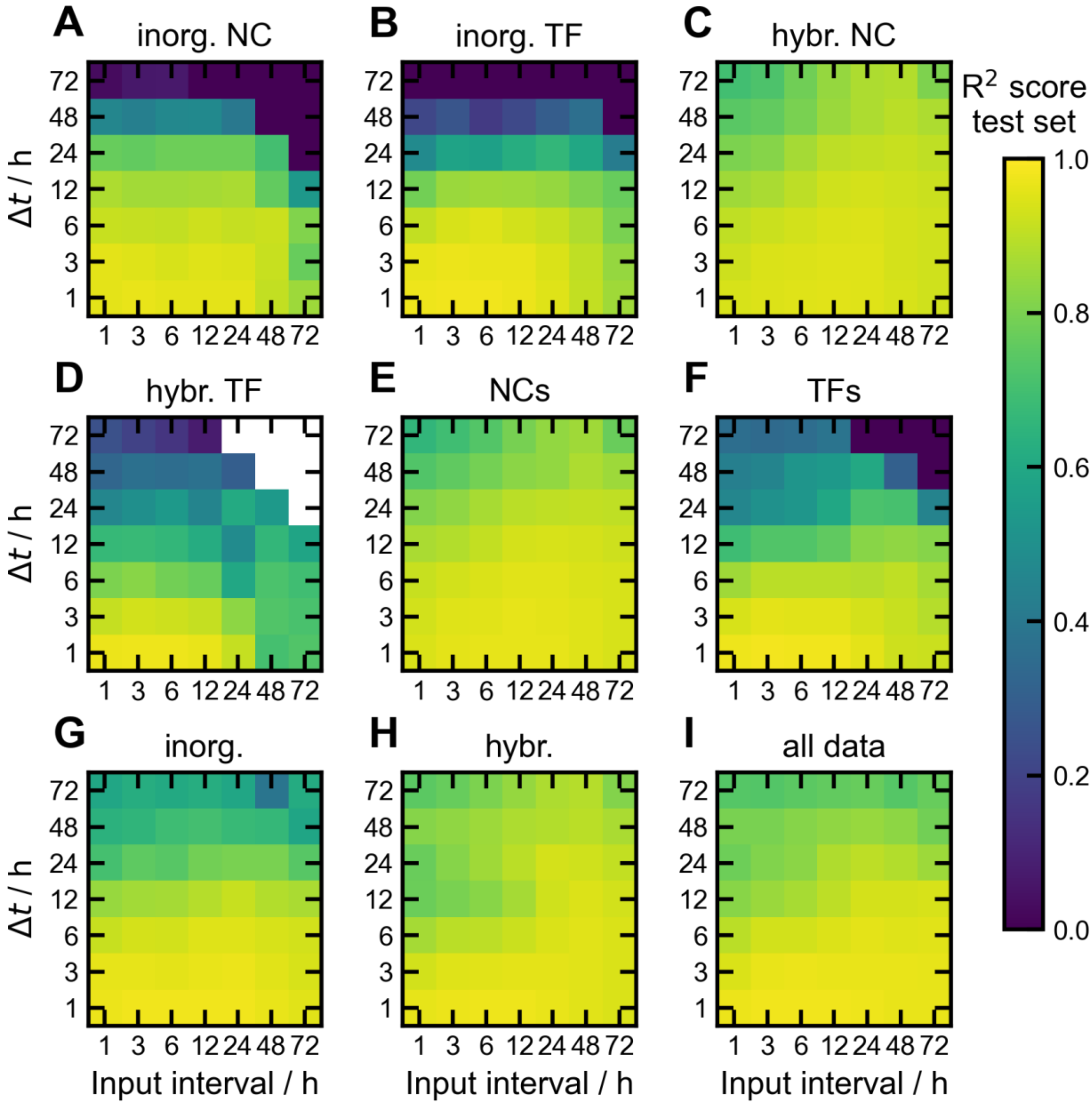

**Figure S13** The coefficient of determination ($R^2$ score) is shown as an average of the 30 train/test/validation splits for the models trained on the individual subsets of the data: **A**: only the inorganic nanocrystals, **B**: only the inorganic thin films, **C**: only the hybrid nanocrystals, **D**: only the hybrid thin films, **E**: both nanocrystal sets, **F**: both thin film sets, **G**: both inorganic sets, **H**: both hybrid sets, and **I**: all the data combined.

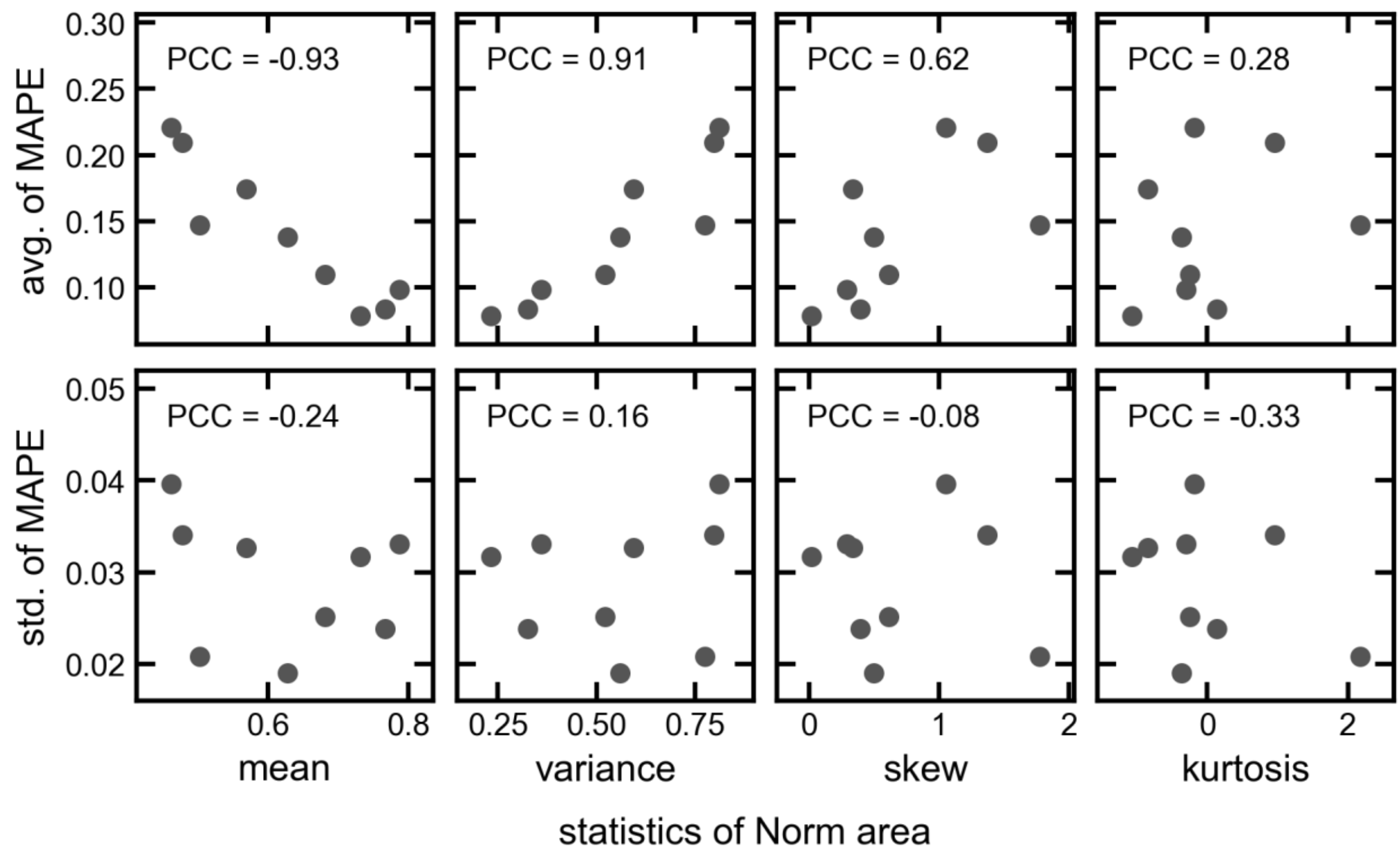

**Figure S14** The statistics of the Norm area for the different datasets in Table S3 are shown together with the average of MAPE and the standard deviation of MAPE across the 30 train/validation/test splits and the 7×7 grids. The Pearson correlation coefficient (PCC) is shown for each plot.

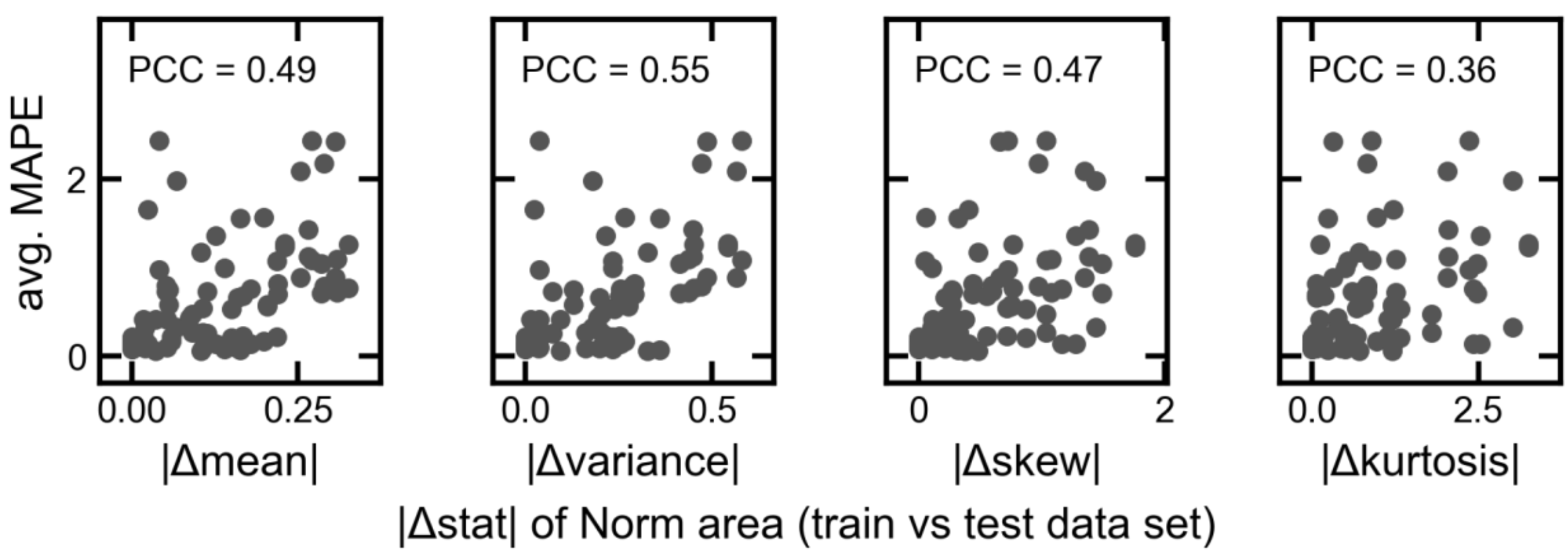

**Figure S15** Differences of the Norm area for the different train and test datasets used in Figure 5 in the main text. The Pearson correlation coefficient (PsCC) is shown for each plot.